\documentclass[twocolumn]{aastex63}
\usepackage{comment}
\usepackage{amsmath}	
\usepackage{amssymb}
\usepackage{footnote}
\usepackage{appendix}
\usepackage[graphicx]{realboxes}
\usepackage{rotating}
\usepackage{threeparttable}

\graphicspath{{./}{figures/}}

\def\ra#1#2#3{#1$^{\rm h}$#2$^{\rm m}$#3$^{\rm s}$}
\def\dec#1#2#3{#1$^\circ$#2$'$#3$''$}

\newcommand\mangoff{\theta}
\newcommand\angoff{$\mangoff$}
\newcommand\mphysoff{\delta R}
\newcommand\physoff{$\mphysoff$}

\newcommand\medangoff{1.1''} 
\newcommand\qangoff{[0.6'',2.9'']} 

\newcommand\medphyoff{3.2~kpc} 
\newcommand\qphyoff{[2.0, 8.7]~kpc} 

\newcommand\medrefoff{1.4} 
\newcommand\qrefoff{[0.6, 2.1]$r_e$} 

\newcommand\mreff{r_e}
\newcommand\reff{$\mreff$}

\newcommand\mssfr{\Sigma_{\rm SFR}} 
\newcommand\ssfr{$\mssfr$}

\newcommand\mmsd{\Sigma_{\rm M*}} 
\newcommand\msd{$\mmsd$}

\newcommand\mFF{F_{\rm F}}  
\newcommand\FF{$\mFF$}
\newcommand\medFF{0.33} 
\newcommand\qFF{[0.20, 0.53]} 

\newcommand\mFE{F_{\rm E}}  
\newcommand\FE{$\mFE$}

\newcommand\mmlim{m_{\rm lim}}
\newcommand\mlim{$\mmlim$}

\newcommand{\galfit}{{\tt GALFIT}}

\def\nod{\nodata}

\shorttitle{{\it HST} FRB Host Galaxies}
\shortauthors{Mannings et al.}

\begin{document}
\sloppy

\title{A High-Resolution View of Fast Radio Burst Host Environments}

\correspondingauthor{A. Mannings}
\email{almannin@ucsc.edu}

\newcommand{\UCSC}{\affiliation{Department of Astronomy and Astrophysics, University of California, Santa Cruz, CA 95064, USA}}

\newcommand{\NU}{\affiliation{Center for Interdisciplinary Exploration and Research in Astrophysics (CIERA) and Department of Physics and Astronomy, Northwestern University, Evanston, IL 60208, USA}}

\newcommand{\IPMU}{\affiliation{Kavli Institute for the Physics and Mathematics of the Universe (Kavli IPMU), 5-1-5 Kashiwanoha, Kashiwa, 277-8583, Japan}}

\newcommand{\STSCI}{\affiliation{
Space Telescope Science Institute, Baltimore, MD 21218, USA
}}

\newcommand{\JHU}{\affiliation{Department of Physics and Astronomy, Johns Hopkins University, Baltimore, MD 21218, USA}}

\newcommand{\iceland}{\affiliation{Centre for Astrophysics and Cosmology, Science Institute, University of Iceland, Dunhagi 5, 107 Reykjav\'ik, Iceland}
}

\newcommand{\Macquarie}{\affiliation{Department of Physics \& Astronomy, Macquarie University, NSW 2109, Australia}
}

\newcommand{\astrophotonics}{\affiliation{Astronomy, Astrophysics and Astrophotonics Research Centre, Macquarie University, Sydney, NSW 2109, Australia}
}

\newcommand{\CSIRO}{\affiliation{Australia Telescope National Facility, CSIRO Astronomy and Space Science, PO Box 76, Epping, NSW 1710, Australia}
}

\newcommand{\Swinburne}{\affiliation{Centre for Astrophysics and Supercomputing, Swinburne University of Technology, Hawthorn, VIC 3122, Australia}}

\newcommand{\tata}{\affiliation{Department of Astronomy and Astrophysics, Tata Institute of Fundamental Research, Mumbai, 400005, India}}

\newcommand{\NCRA}{\affiliation{National Centre for Radio Astrophysics, Post Bag 3, Ganeshkhind, Pune, 411007, India}}

\newcommand{\PUCV}{\affiliation{Instituto de F\'isica, Pontificia Universidad Cat\'olica de Valpara\'iso, Casilla 4059, Valpara\'iso, Chile}}
\author{Alexandra G. Mannings}
\UCSC

\author[0000-0002-7374-935X]{Wen-fai Fong}
\NU

\author[0000-0003-3801-1496]{Sunil Simha}
\UCSC

\author[0000-0002-7738-6875]{J.~Xavier~Prochaska}
\UCSC
\IPMU

\author[0000-0002-9946-4731]{Marc~Rafelski}
\STSCI
\JHU

\author{Charles D. Kilpatrick}
\NU

\author[0000-0002-1883-4252]{Nicolas Tejos}
\PUCV

\author[0000-0002-9389-7413]{Kasper E. Heintz}
\iceland

\author{Keith W. Bannister}
\CSIRO

\author[0000-0003-3460-506X]{Shivani Bhandari}
\CSIRO

\author{Cherie K. Day}
\Swinburne
\CSIRO

\author{Adam T. Deller}
\Swinburne

\author[0000-0003-4501-8100]{Stuart D. Ryder}
\Macquarie
\astrophotonics

\author{Ryan M. Shannon}
\Swinburne

\author[0000-0003-2548-2926]{Shriharsh P. Tendulkar}
\tata
\NCRA

\begin{abstract}
    We present {\it Hubble Space Telescope} ({\it HST}/WFC3) ultraviolet and infrared observations of eight fast radio burst (FRB) host galaxies with sub-arcsecond localizations, including the hosts of three known repeating FRBs. We quantify their spatial distributions and locations with respect to their host galaxy light distributions, finding that they occur at moderate host normalized-offsets of \medrefoff $r_e$ (\qrefoff; 68\% interval), occur on fainter regions of their hosts in terms of IR light, but overall trace the radial distribution of IR light in their galaxies. The FRBs in our tested distribution do not clearly trace the distributions of any other transient population with known progenitors, and are statistically distinct from the locations of LGRBs, H-poor SLSNe, SGRBs, and Ca-rich transients. We further find that most FRBs are not in regions of elevated local star formation rate and stellar mass surface densities in comparison to the mean global values of their hosts. 
    We also place upper limits to the IR flux
    at the FRB positions of $m_{\rm IR}\gtrsim\!24.8-27.6$~AB~mag, constraining both satellite and background galaxies to luminosities well below the host luminosity of FRB\,121102. We find that 5/8 FRB hosts exhibit clear spiral arm features in IR light, and that the positions of all well-localized FRBs located in such hosts are consistent with their spiral arms, although not on their brightest regions. Our results do not strongly support the primary progenitor channel of FRBs being connected either with the most massive (stripped-envelope) stars, or with events which require kicks and long delay times (neutron star mergers).
\end{abstract}

\keywords{radio transient sources -- Hubble Space Telescope}

\section{Introduction}

Since the discovery of fast radio bursts (FRBs) in 2007 \citep{Lorimer07} -- dispersed, millisecond-timescale transients, traced primarily to cosmological distances \citep{Thornton13,CordesChatterjee19} -- 
their definitive origins have remained elusive. The emerging association of FRBs with magnetized neutron stars (``magnetars'') was catalyzed by the discovery and sub-arcsecond localization of the repeating FRB\,121102 \citep{Spitler16,Chatterjee17,Tendulkar17}, which was found to be spatially consistent with a compact, persistent radio source \citep{Chatterjee17} postulated to be a synchrotron nebula powered by an embedded magnetar \citep{Marcote17,Margalit18} or an offset active galactic nucleus (AGN; \citealt{Marcote17,Eftekhari20}). The association of at least some FRBs with magnetars was further solidified by the detection of multiple FRB events from the Galactic magnetar SGR\,1935+2154 \citep{SGRDiscovery2020,Bochenek20}. However, a myriad of progenitor models still exist \citep{Platts19,Petroff19}, and the precise connection to magnetars for the extragalactic population has yet to be made definitive. Moreover, given the diversity of their observed FRB properties, it is not clear if there is one or multiple progenitor channels for FRB production.

The advent of dedicated FRB experiments are fueling an unprecedented rise in their detection rates \citep{CHIME18,Macquart10}. One of the most promising ways to make progress on their origins is through robust associations to host galaxies, which generally requires precise localizations of $\lesssim\!1''$ \citep{Eftekhari17}. Indeed, different progenitor channels will yield distinct host galaxy demographics and host stellar population properties (e.g., \citealt{Margalit19}). It is also expected that different production pathways will be imprinted in how FRBs are distributed with respect to their host galaxies. Prior to recent advancements in our ability to localize FRBs, surveys were only able to produce arc-second localizations for repeating bursts such as 121102 and 180916. Repeating bursts, however, are thought to only make up $ \simeq 2-5 \%$ of all bursts. This means that, as of now, repeating bursts are over-represented in the sample of precisely localized FRBs. With surveys such as CRAFT \citep{Shannon18} coming online which provide sub-arcsecond localizations of apparently non-repeating bursts, we can form a more complete picture of FRB host characteristics and therefore production pathways.

Locations have historically played an important role in delineating the progenitors of a wide range of transients. For instance, the spatial distributions and strong correlation with the UV light distributions of their hosts for super-luminous supernovae (SLSNe) and long-duration gamma-ray bursts (LGRBs) is commensurate with their massive star origins \citep{Woosley93,Fruchter06,Lunnan15,Perley16}, while the significant offsets of short-duration gamma-ray bursts (SGRBs) and weaker correlation to UV light is indicative of an older stellar progenitor \citep[e.g.][]{Fong13}. Moreover, quantifying the relationship between the locations of core-collapse SNe (CCSNe, Types Ib/c, II) and Type Ia SNe, to detailed morphological features such as spiral arms, can serve as an indirect indicator of the age of their stellar and/or white dwarf progenitors \citep{Audcent-Ross20}.

For FRBs, two primary pathways that have been considered for magnetar formation are ``prompt'' magnetars, formed from massive star progenitors, and ``delayed''-channel magnetars, formed from existing compact objects and their interactions, e.g., neutron star mergers or accretion-induced collapse (AIC) of a white dwarf to a NS \citep{Nicholl17,Margalit19}. Detailed studies using ground-based observations based on the first $\sim$dozen well-localized FRBs and their host galaxies have shown that their spatial distributions are inconsistent with engine-driven massive star explosions (LGRBs, SLSNe; \citealt{Li19,Bhandari20a,Heintz20, Bochenek20}).

One is therefore motivated to characterize, as precisely
as possible, the local environments of FRBs
within their host galaxies.  Furthermore, the competing progenitor models offer distinct predictions
for the ages and masses of the responsible compact
object(s).  For young progenitors, there must be 
a direct link to ongoing or recent star-formation 
activity.  Older progenitor channels, meanwhile, may
track the underlying stellar mass of the galaxy.
Thus motivated, we have designed an experiment 
to examine the local environments of FRBs, in both
active star-formation (via near-UV light)
and stellar mass (near-IR) and at the highest
spatial resolution afforded by space-borne
instrumentation.

Against this backdrop, we present the first comprehensive sample of {\it HST} observations for eight FRB host galaxies, six of which are newly presented in this work. In Section~\ref{sec:obs}, we describe our sample, observations and data reduction. In Section~\ref{sec:offsets}, we present the spatial distribution of FRBs (angular, physical and host-normalized offsets) and the locations of FRBs with respect to their host light distributions. In Section~\ref{sec:explosion_sites}, we present the results of surface brightness profile fits, including the revelation of spiral structure in several hosts; constraints on the star formation rate and stellar mass densities at the FRB positions; and deep limits on possible satellite or background galaxies. In Section~\ref{sec:discuss}, we discuss the implications of our results in terms of FRB progenitors. We highlight our main conclusions in Section~\ref{sec:conc}. Throughout the paper, we employ a Planck cosmology with $H_0 = 67.8$~km~s$^{-1}$~Mpc$^{-1}$, $\Omega_M = 0.308$, and $\Omega_{\Lambda}=0.692$ \citep{Planck15}.  All of the data and analysis code are made available via GitHub\footnote{https://github.com/FRBs/FRB}.


\begin{deluxetable*}{lcccccllcccc}[t!]
\tabletypesize{\footnotesize}
\tablecaption{Log of {\it Hubble Space Telescope} FRB Host Galaxy observations \label{tab:obs}}
\tablecolumns{9}
\tablewidth{0pt}
\tablehead{
\colhead{FRB} &
\colhead{RA$_{\rm Host}$} & 
\colhead{Dec$_{\rm Host}$} & 
\colhead{$\sigma_{\rm Host}$} &
\colhead{$z$} & 
\colhead{$M_{*}$} &
\colhead{Instrument} & 
\colhead{Filter} & 
\colhead{Date} & 
\colhead{Exp. Time}
\\
\colhead{} & 
\colhead{(J2000)} &
\colhead{(J2000)} & 
\colhead{(mas)} & 
\colhead{} & 
\colhead{($10^{9} M_{\odot}$)} &
\colhead{} & 
\colhead{} & 
\colhead{(UT)} & 
\colhead{(sec)}
}
\startdata
121102 & \ra{05}{31}{58.69} & $+$\dec{33}{08}{52.43} & 6.3 & 0.1927 & $0.14 \pm 0.07$ & WFC3/IR & F160W & 2017 Feb 23 & 1197 \\ 
       &                   &                        &     &        &                  & WFC3/UVIS & F763M & 2017 Feb 23 & 1940 \\
180916 & \ra{01}{58}{00.29} & $+$\dec{65}{42}{53.09} & 1.8 & 0.0337 & $2.15 \pm 0.33$ & WFC3/IR & F110W & 2020 Jul 17 & 5623  \\
       &                   &                        &     &        &                  & WFC3/UVIS & F673N & 2020 Jul 16 & 2877 \\
180924 & \ra{21}{44}{25.256} & $-$\dec{40}{54}{00.80} & 0.4 & 0.3212 & $13.23 \pm 5.06$ & WFC3/IR & F160W & 2019 Nov 27 & 2470 \\
       &                     &                        & 3.1 &      &     & WFC3/UVIS & F300X & 2019 Nov 26 & 2492 \\
190102 & \ra{21}{29}{39.577} & $-$\dec{79}{28}{32.52} & 14.2 & 0.2912 & $3.39 \pm 1.02$ & WFC3/IR & F160W& 2020 Jan 14 & 2470\\
       &                     &                        & \nod$^{\dagger}$ &  & & WFC3/UVIS& F300X&2019 Oct 07&2776\\
190608 & \ra{22}{16}{04.903} & $-$\dec{07}{53}{55.91} & 0.8 & 0.1177 & $11.57 \pm 0.84$ & WFC3/IR & F160W & 2019 Dec 01 & 2295\\
       &                     &                       & 0.8 &       &    &    WFC3/UVIS& F300X& 2019 Oct 11& 2400\\
190711 & \ra{21}{57}{40.613} & $-$\dec{80}{21}{29.05} & 5.2 & 0.522 & $0.81 \pm  0.29$  & WFC3/IR & F160W & 2020 May 11 & 2470\\
       &                     &                        & 7.7 &      &     &   WFC3/UVIS & F300X & 2020 May 09 & 2780\\
190714 & \ra{12}{15}{55.090} & $-$\dec{13}{01}{15.96} & 1.1 & 0.2365 & $14.92 \pm 7.06$ & WFC3/IR & F160W & 2020 Apr 30 & 2295\\
       &                     &                        & \nod$^{\dagger}$ & & & WFC3/UVIS & F300X & 2020 May 19 &  2396\\
191001 & \ra{21}{33}{24.440} & $-$\dec{54}{44}{54.53} & 0.5 & 0.2340 & $46.45 \pm 18.80$ & WFC3/IR & F160W & 2020 Apr 28 & 2296 \\
       &                     &                        & \nod$^{\dagger}$ & & & WFC3/UVIS & F300X & 2020 Apr 25 & 2580
\enddata
\tablecomments{Data are from programs 15878 (FRB\,180924, 190102, 190608, PI:~Prochaska), 16080 (FRBs\,190711, 190714, and 191001, PI:~Mannings), 14890 (FRB\,121102, PI:~Tendulkar), and 16072 (FRB\,180916, PI:~Tendulkar). \\
$\sigma_{\rm Host}$ is the $1\sigma$ positional uncertainty of the host (RA and Dec components added in quadrature). $M_{*}$ is the host stellar mass. \\
$^{\dagger}$ S/N of host galaxy is not sufficient to obtain an uncertainty on the position. \\
{\bf Redshift References--} \citet{Tendulkar17,Bannister19,Marcote20,Bhandari20a,Heintz20} }
\end{deluxetable*}


\section{Data}
\label{sec:obs}
\subsection{Sample of FRB Host Galaxies}
Here we present observations of eight FRB host galaxies obtained with the Wide-Field Camera~3 using the infrared and ultra-violet-visual channels (WFC3/IR and WFC3/UVIS). The data for six of the FRB host galaxies were collected between October 2019 and April 2020 as part of programs 15878 (PI: Prochaska) and 16080 (PI: Mannings), which targeted galaxies for which FRB events have been detected and localized by the Commensal Real-time ASKAP Fast Transients (CRAFT) survey on the Australian Square Kilometer Array Pathfinder (ASKAP; \citealt{Bannister19, Day20, Bhandari20a, Chittidi20, Macquart20}). These bursts were localized to sub-arcsecond precision, with $\sigma_{\rm FRB} \approx 0.1-0.7''$. 

We also include two additional FRB hosts with {\it HST} observations, FRB\,121102\footnote{The Transient Name Server (TNS) name for this burst is FRB\,20121102a.} \citep{Bassa17} under program 14890 (PI:~Tendulkar) taken in February 2017 and FRB\,180916 \citep{tendulkar20} under program 16072 (PI:~Tendulkar) taken in July 2020. FRB\,121102 was discovered by the Arecibo telescope \citep{Spitler16}, and subsequently localized via repeating bursts with the Very Large Array (VLA; \citealt{Chatterjee17}), and with Very Long Baseline Interferometry (EVN VLBI; \citealt{Marcote17}) with $\sigma_{\rm FRB}=0.0045''$. FRB\,180916 is the closest and most precisely localized FRB with $\sigma_{\rm FRB} = 0.0023''$ \citep{Marcote20}. Our sample thus comprises all eight FRB host galaxies for which there exist available {\it HST} observations.
Table~\ref{tab:obs} summarizes all of these data
and Table~\ref{tab:frbs} summarizes coordinates
and the localization errors of the FRBs.

\begin{deluxetable*}{cccccccc}
\tablewidth{0pc}
\tablecaption{FRB Sample and Localizations\label{tab:frbs}}
\tabletypesize{\normalsize}
\tablehead{\colhead{FRB}  
& \colhead{RA$_{\rm FRB}$} & \colhead{Dec$_{\rm FRB}$}  
& \colhead{$a_{\rm stat}$} & \colhead{$a_{\rm sys}$} 
& \colhead{$b_{\rm stat}$} & \colhead{$b_{\rm sys}$} 
& \colhead{PA} 
\\& (J2000) & (J2000) &  ($''$) & ($''$) & ($''$) & ($''$) & (deg)  
} 
\startdata 
121102& 82.994589& $33.1479316$& 0.004& 0.00& 0.002& 0.00& 90.0\\ 
180916& 29.503126& $65.7167542$& 0.001& 0.00& 0.001& 0.00& 0.0\\ 
180924& 326.105229& $-40.9000278$& 0.07& 0.09& 0.06& 0.07& 0.0\\ 
190102& 322.415667& $-79.4756944$& 0.21& 0.52& 0.17& 0.44& 0.0\\ 
190608& 334.019875& $-7.8982500$& 0.19& 0.19& 0.18& 0.18& 90.0\\ 
190611& 320.745458& $-79.3975833$& 0.34& 0.60& 0.32& 0.60& 0.0\\ 
190711& 329.419500& $-80.3580000$& 0.12& 0.38& 0.07& 0.32& 90.0\\ 
190714& 183.979667& $-13.0210278$& 0.17& 0.32& 0.10& 0.23& 90.0\\ 
191001& 323.351554& $-54.7477389$& 0.13& 0.11& 0.08& 0.10& 90.0\\ 
200430& 229.706417& $12.3768889$& 0.01& 0.02& 0.24& 1.00& 0.0\\ 
\hline 
\enddata 
\tablecomments{$a_{\rm stat}, a_{\rm sys}$ are the angular size of the semi-major axis describing the $1\sigma$ statistical and systematic uncertainties respectively.
$b$ refers to the semi-minor axis.  PA is the sky position angle of the error ellipse.
Sources without reported systematic error have been incorporated in the statistical.
Data from \cite{Day20,Tendulkar17,Marcote20,Heintz20}.}
\end{deluxetable*}

All of the host galaxies in our {\it HST} sample have spectroscopically-confirmed redshifts. These are considered secure associations\footnote{See also \cite{PATH21}
for a Bayesian analysis that reaches
similar conclusions.} 
with probabilities of chance coincidence of $P_{\rm chance} \lesssim$ 0.05 \citep{Heintz20} with their most likely host galaxy. The {\it HST} data for FRB\,121102 and FRB\,180916 were previously published in \citet{Bassa17} and \citet{tendulkar20}, respectively, while the WFC3/UVIS image for FRB\,190608 and its local environment was previously published and analyzed in \citet{Chittidi20}. All of the remaining {\it HST} observations are newly presented here. Three of the bursts are known ``repeating'' FRBs (FRBs\,121102, 190711 and 180916; \citealt{Spitler16,Kumar20}) while the remaining bursts are apparent ``non-repeaters''.

We supplement this sample with ground-based data from two other FRB hosts presented in \citet{Heintz20} with secure host associations (FRB\,190611 and FRB\,200430) when computing cumulative distributions of offsets in Section ~\ref{ssec:off}. Both of these FRBs in the ground-based sample are apparent non-repeaters. Combined, our ground-based and {\it HST} sample comprises all FRBs with sub-arcsecond localizations discovered over 2012-2020. The exception is FRB\,190614D which does not have a clear host galaxy association \citep{Law20},
and is not included in the sample.

In current surveys (including this one), repeating bursts are over-represented within the sample of precisely localized FRBs. Because of their repetition, they are more likely to be localized and make up around half of these precisely localized bursts. Only ~5\% of FRBs are known to repeat ((citation)). differentiating between repeaters and non-repeaters will be important in future studies with larger sample sizes, as the non-repeaters will represent the majority of all FRBs.

\subsection{Observations}

For the WFC3/UVIS observations under programs 15878 and 16080, we use the ultra-wide F300X filter to sample the rest-frame near- and far-ultraviolet (NUV/FUV) wavelengths with the aim of capturing the distribution of star formation in the host galaxies. This filter provides increased throughput in the NUV compared to the standard wide filters (although it has a minor red tail out to $\sim$4000~\AA), and is chosen to maximize the signal-to-noise (S/N) in a single orbit of {\it HST} imaging. To minimize the effects of charge transfer efficiency (CTE) degradation, we set up the observations to position the targets near the readout on amplifier C located on chip 2. We used a 4-point dither pattern to sub-sample the point spread function (PSF) and remove detector artifacts. We increase the line and point spacing by a factor of 5 over the standard box pattern to remove residual background patterns as described in \citet{Rafelski15}. The data from program 15878 include a 9~e$^-$ post-flash per exposure to reach 12~e$^-$ per pixel background. Recently there was a new recommendation to reach a background of 20e$^-$, and therefore the data obtained in program 16080 included a 17~e$^-$ post-flash to reach this level \citep{Anderson12}.

For the WFC3/IR observations under programs 15878 and 16080, we use the F160W filter, the reddest wide filter available with {\it HST}, to cover the rest-frame optical band to assess the distribution of the stellar mass as traced by older stellar populations. We use SPARS25 and NSAMP15 to ensure that the observations remain in the linear count regime, and obtain clean images by dithering over the known IR ``blobs'' with a seven-point dither pattern with a factor 3 increase in spacing of the 7 point wide dither pattern provided in \citet{Anderson16}. Finally, the data for FRB\,121102 under program 14890 employ a two-point and four-point dither pattern for the WFC3/UVIS and IR observations, while the data for 180916 employ a three point and four-point dither pattern, respectively \citep{Bassa17,tendulkar20}. The details of these observations are listed in Table~\ref{tab:obs}.

\subsection{Image Processing}
The data were retrieved from the Barbara A. Mikulski Archive for Space Telescopes (MAST), and the WFC3/UVIS data are custom calibrated. These data have degraded CTE, and therefore require pixel-based CTE corrections \citep{Anderson12}.  In addition, the degradation requires improved dark, hot pixel, and bias level corrections. First, we use a new correction for the CTE. Second, we apply concurrent superdarks to the data, reducing the blotchy pattern otherwise present \citep{Rafelski15}. Third, we identify hot pixels in the darks such that the number of hot pixels is consistent as a function of the distance to the readout amplifiers based on the number of hot pixels identified close to the readout. This is accomplished by modifying the threshold for hot pixel detection as a function of distance to the readout (Prichard et al. in prep). Lastly, we normalize the amplifiers since the applied superbias is based on bias files with insufficient background levels for a pixel based CTE correction. We measure the background level in each amplifier after masking sources, and match the background level between the amplifiers. 

\begin{figure*}
    \centering
    \includegraphics[width=0.8\textwidth,trim={0 0.2in 0 0.75in},clip]{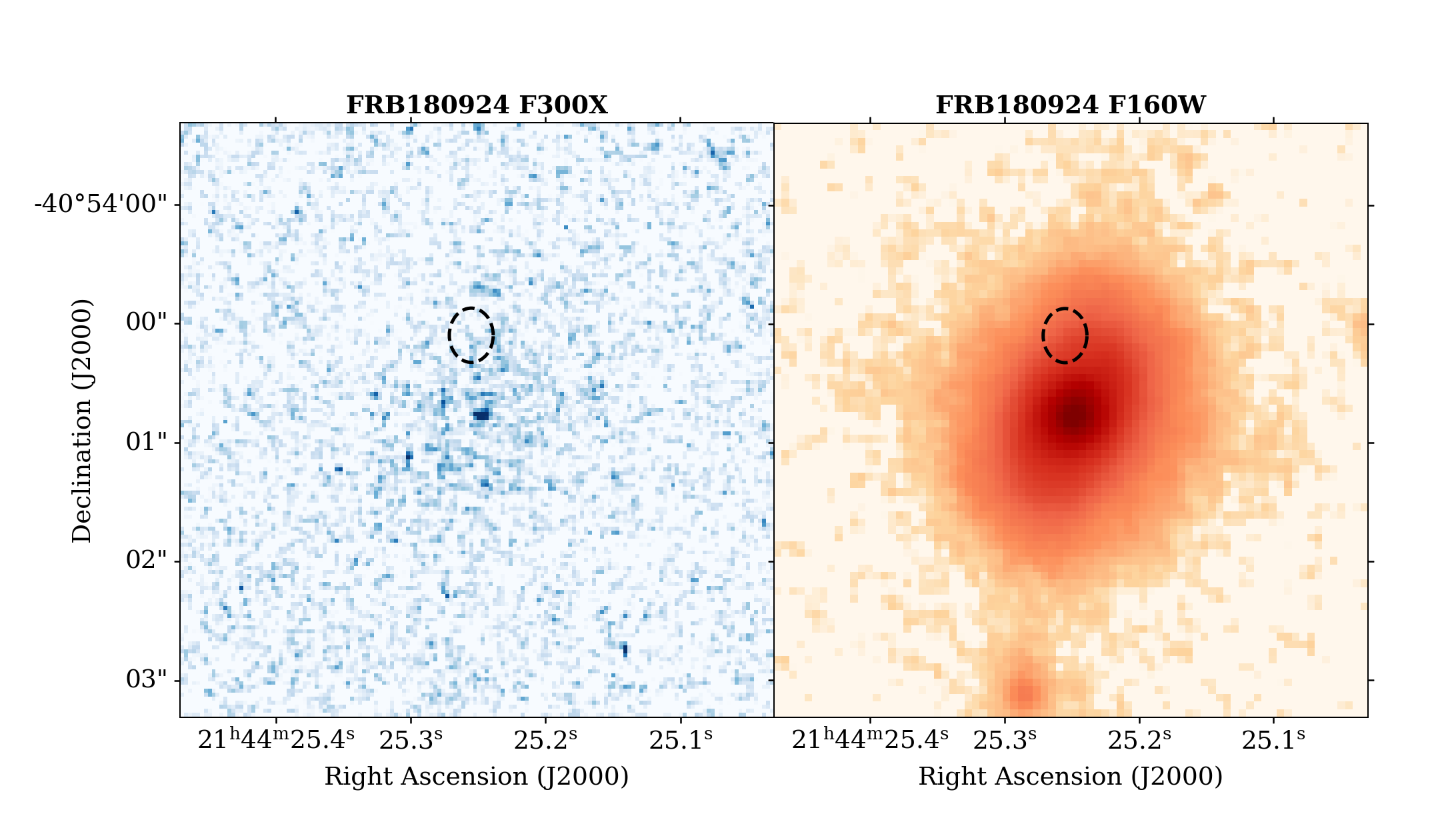} \\
    \includegraphics[width=0.8\textwidth,trim={0 0.2in 0 0.75in},clip]{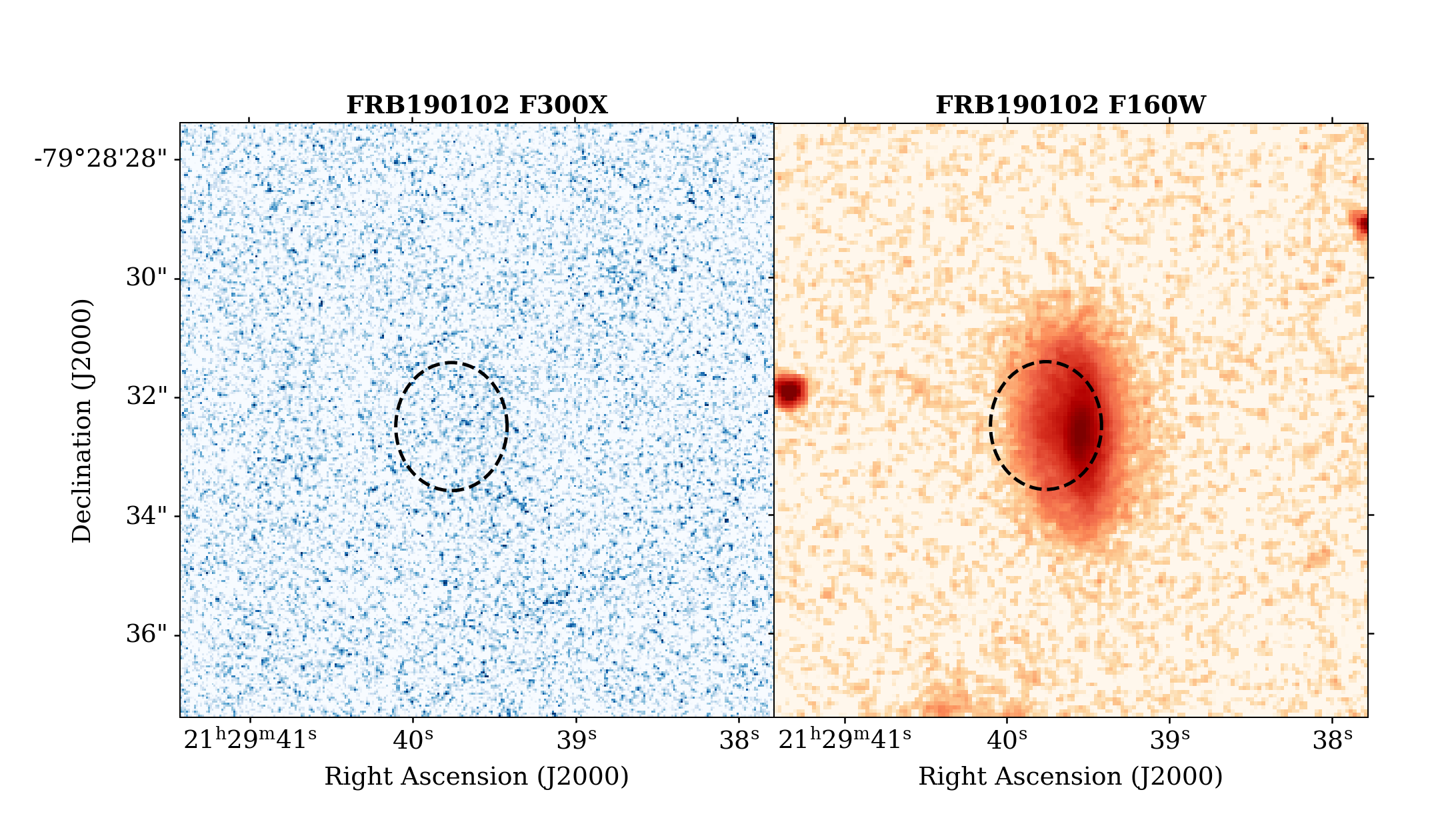} \\
    \includegraphics[width=0.8\textwidth,trim={0 0.2in 0 0.75in},clip]{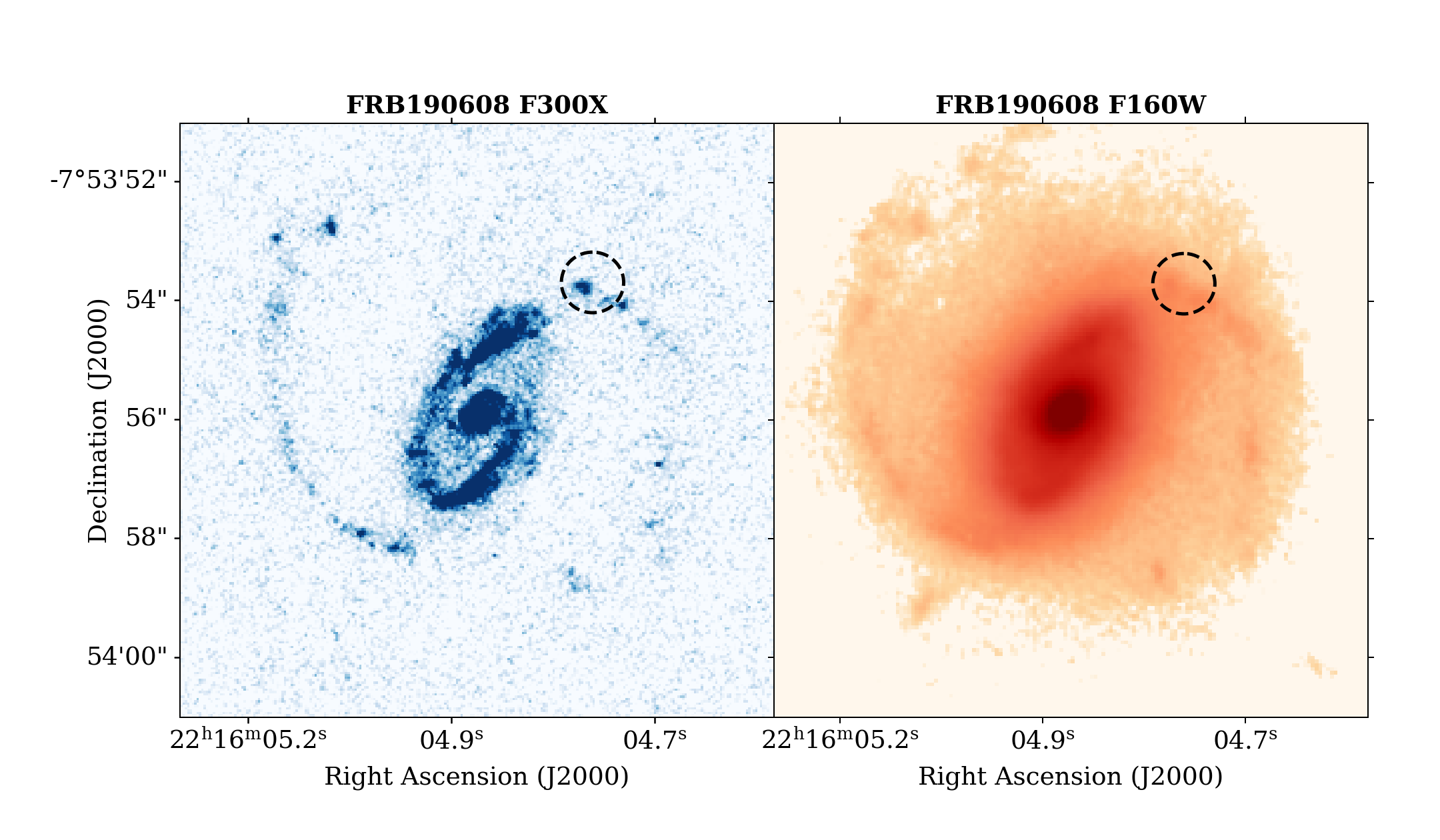} \\
    \caption{{\it HST} imaging of three of the host galaxies in our sample, for FRB\,180924, FRB\,190102, and FRB\,190106. Images with a blue color-bar were taken with the UV channel F300X filter, while images in the red color-bar were taken with the IR channel F160W filter. The black ellipse in each image denotes the FRB position ($2\sigma$ uncertainty in each coordinate). All images are oriented with North up and East to the left.}
    \label{fig:frb_images1}
\end{figure*}

\begin{figure*}
    \centering
    \includegraphics[width=0.8\textwidth,trim={0 0.2in 0 0.75in},clip]{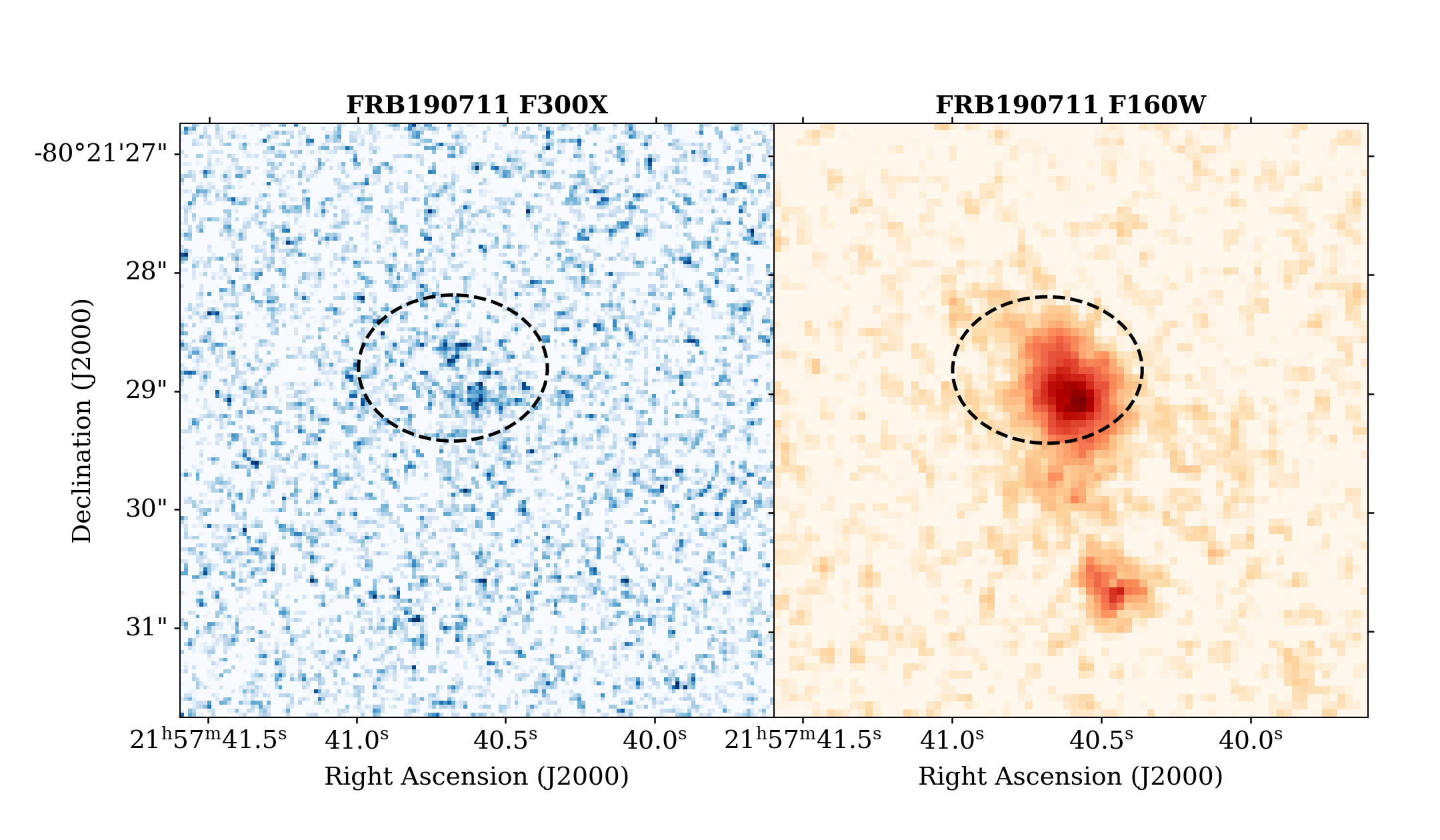} \\
    \includegraphics[width=0.8\textwidth,trim={0 0.2in 0 0.75in},clip]{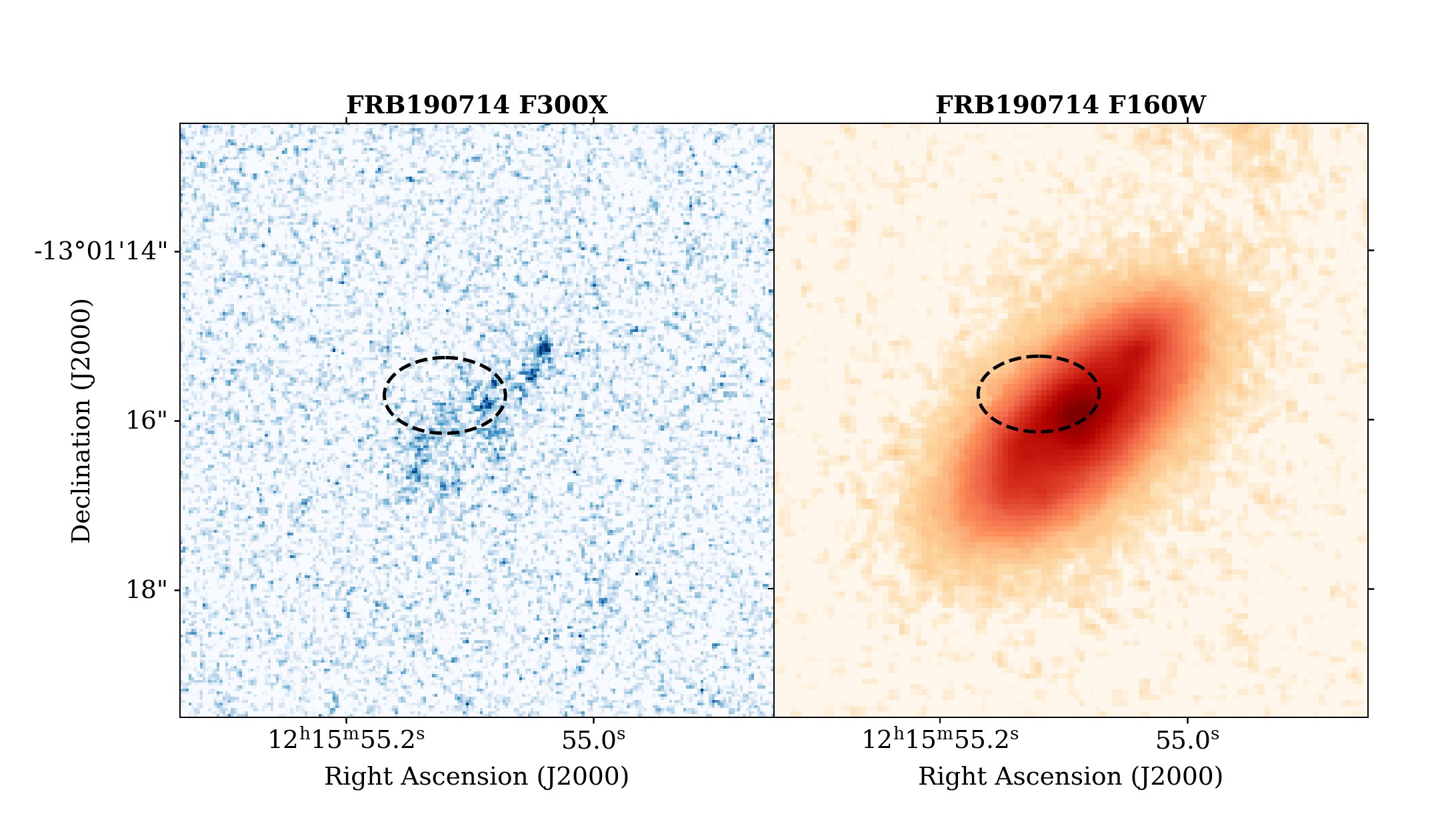} \\
    \includegraphics[width=0.8\textwidth,trim={0 0.2in 0 0.75in},clip]{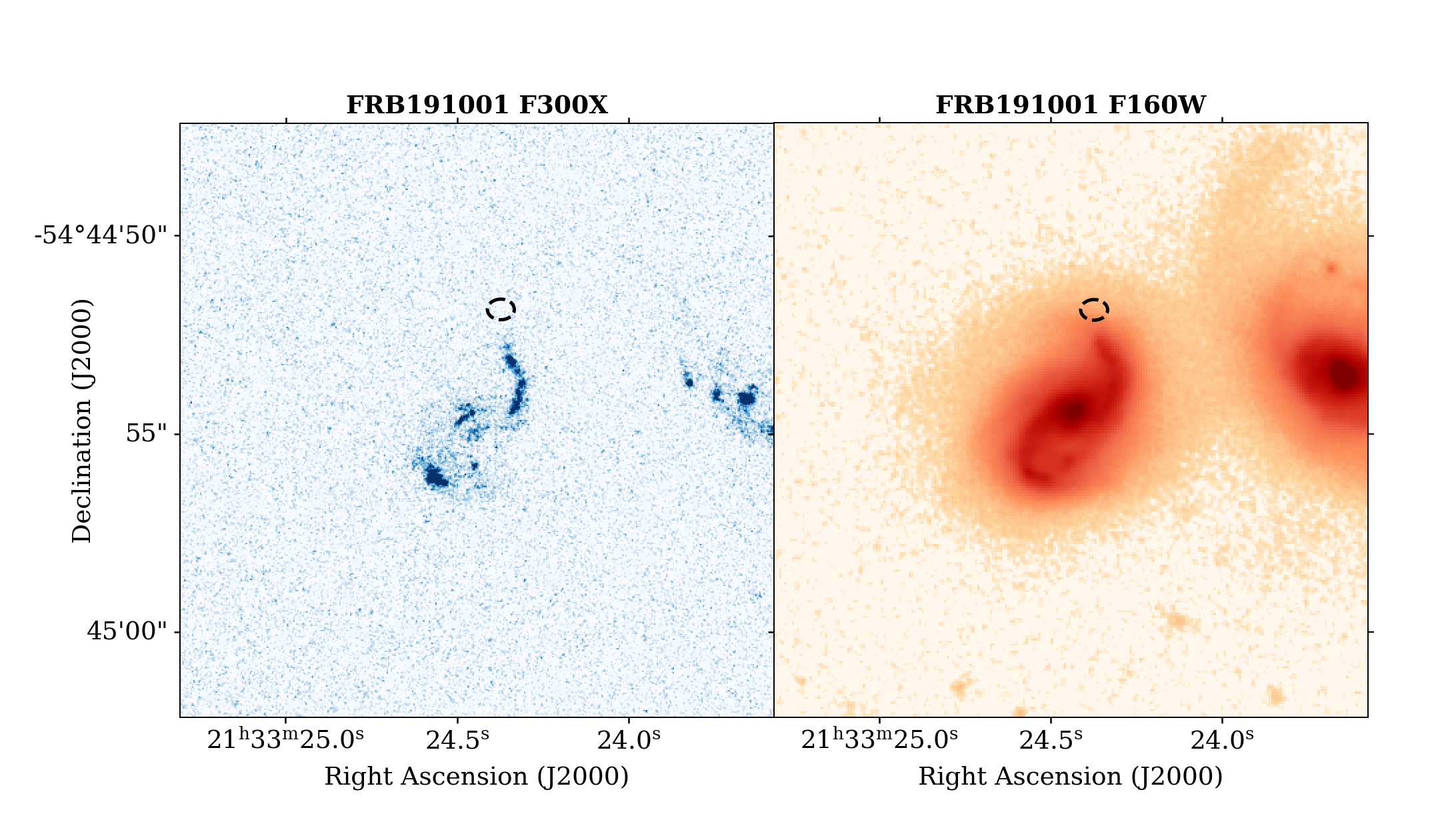} \\
    \caption{{\it HST} imaging of three of the host galaxies in our sample, for FRB\,190711, FRB\,190714, and FRB\,191001. Color scheme and ellipses are as in Figure~\ref{fig:frb_images2}.}
    \label{fig:frb_images2}
\end{figure*}

\begin{figure*}
    \centering
    \includegraphics[width=0.42\textwidth,trim={0 0.1in 0 0.6in},clip]{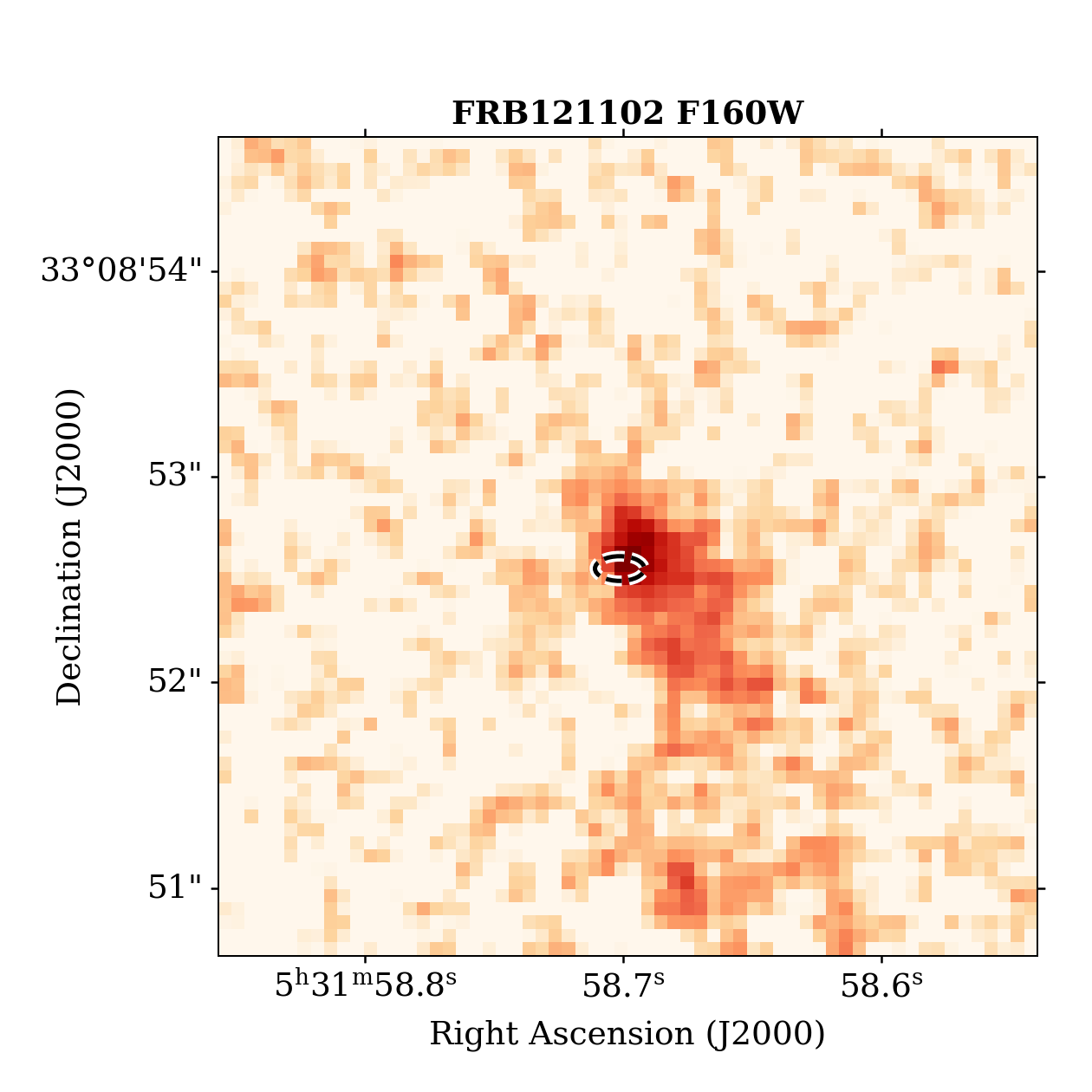}
    \includegraphics[width=0.42\textwidth,trim={0 0.1in 0 0.6in},clip]{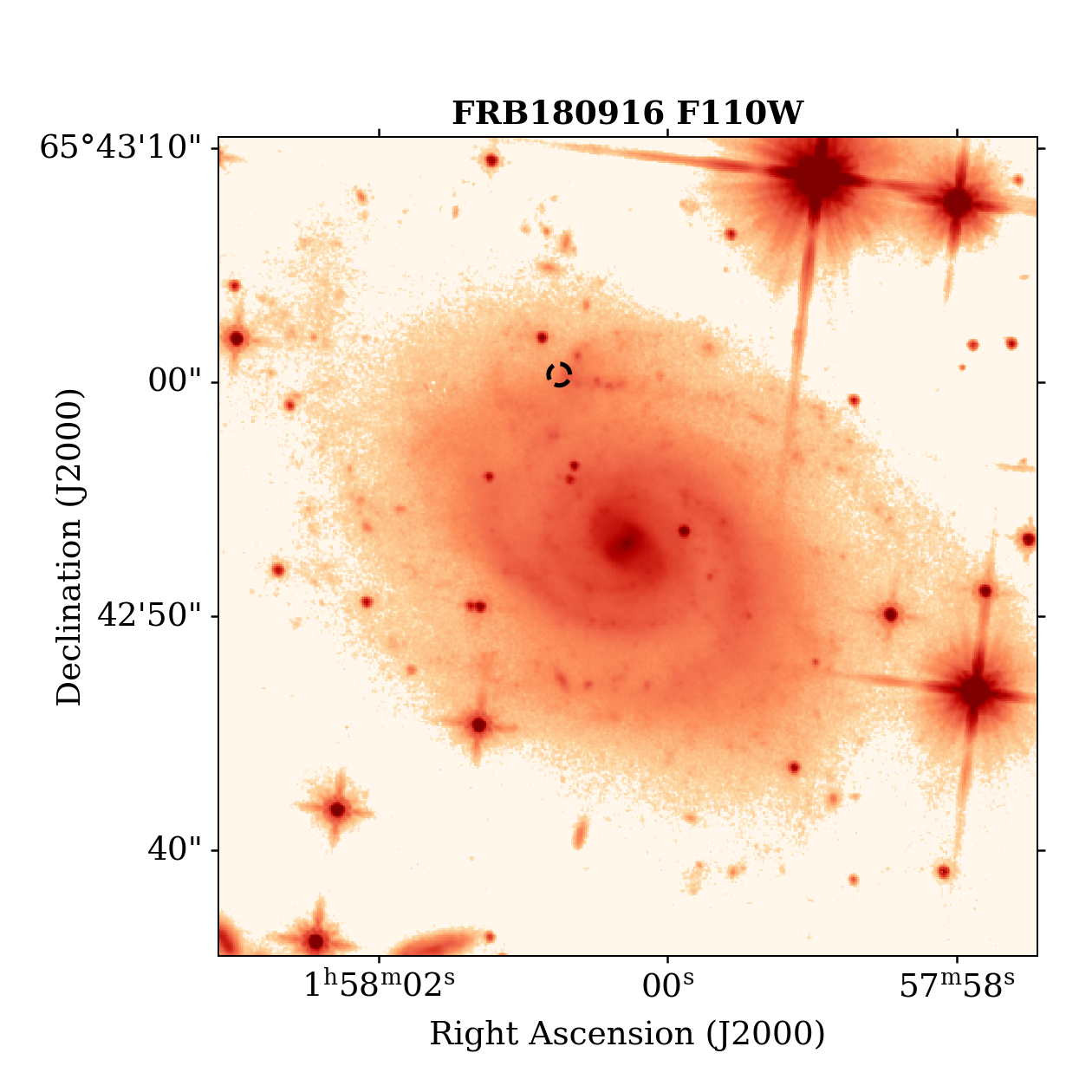}
    \vspace{-0.1in}
    \caption{{\it HST} IR imaging of the host galaxies of FRB\,121102 (F160W) and FRB\,180916 (F110W). The black dashed ellipse represents 30 times the $1\sigma$ uncertainty from the VLBI localization for FRB\,121102 \citep{Marcote17} and 200 times the $1\sigma$ uncertainty for FRB\,180916 \citep{Marcote20}.}    \label{fig:frb_images3}
\end{figure*}

To combine the images for each FRB and in each filter, we used the AstroDrizzle routine as part of the DrizzlePac software package \citep{Avila15} employing a \texttt{pixfrac}$ =0.8$, \texttt{pixscale}$ =0.033''$ for UVIS images, and $0.064''$ for IR images. As part of AstroDrizzle, we also perform cosmic ray removal and sky subtraction. The final drizzled images are shown in Figures~\ref{fig:frb_images1}-\ref{fig:frb_images3}.

\section{FRB Locations \& Offsets}
\label{sec:offsets}

In this section, we present the locations of the FRBs in our sample with respect to their host galaxy centers. 
We introduce their angular and physical offsets 
(\angoff\ and \physoff, respectively), 
``host-normalized'' offsets ($\mphysoff/\mreff$) 
which are normalized by the half-light radii \reff\ of their host galaxies, the location with respect to their host galaxy light distribution (``fractional flux''; \FF), and the 
fraction of light enclosed \FE\ within the radius of the FRB.

\subsection{Astrometry \& Uncertainties}
\label{sec:astrometry}

We first perform absolute astrometry using sources in common between available deep, optical ground-based imaging and the Gaia~DR2 catalog. The ground-based imaging is sourced from the Gemini-North Observatory (FRB\,121102), Gemini-South Observatory (FRB\,190711), Very Large Telescope (VLT; FRBs\,190102, 190714, 191001), the Dark Energy Survey (DES; FRB\,180924) and the Sloan Digital Sky Survey (SDSS; FRB\,190608). We then perform relative astrometry to tie the ground-based images to the WFC3/IR images. This series of astrometric ties ensures that there are enough sources in common with the WFC3 imaging, which has a significantly narrower field-of-view than the ground-based imaging, to properly calculate the total astrometric uncertainty. For astrometry, we employ a six-order polynomial accounting for linear shifts, rotation, and skew, using IRAF/{\tt ccmap} and {\tt ccsetwcs}. We calculate a range of tie uncertainties in each coordinate, of $\sigma_{\rm tie, RA} \approx 0.014-0.073''$ and $\sigma_{\rm tie, Dec} \approx 0.015-0.097''$.

We also use the $\texttt{SourceExtractor}$ tool \citep{SExtractor} to determine the FRB galaxy centroid positions and associated uncertainties ($\sigma_{\rm host}$). These positions and values for $\sigma_{\rm host}$ are listed in Table~\ref{tab:obs}. The final source of uncertainty is the positional uncertainty of the FRB ($\sigma_{\rm FRB}$), which is derived from the statistical and systematic uncertainties from the FRB detections
(Table~\ref{tab:frbs}).


\begin{deluxetable*}{cccccccccccccccc}
\tablewidth{0pc}
\tablecaption{Offsets and Light Locations of FRBs\label{tab:locations}}
\tabletypesize{\normalsize}
\tablehead{\colhead{FRB}  
& \colhead{Filter} & \colhead{\angoff} & \colhead{\physoff} 
& \colhead{\physoff/\reff} & \colhead{\FF} & \colhead{\FE} 
\\&  &  ($''$) & (kpc) &  &  
} 
\startdata 
121102& F160W& 0.23 $\pm$ 0.02& 0.75$ \pm$ 0.05& 0.37 $ \pm$ 0.02& 0.70 $\pm$ 0.07&0.24 $\pm$ 0.12\\ 
& F763M&&&& 0.67 $\pm$ 0.10&\\ 
180916& F110W& 7.760 $\pm$ 0.023& 5.386$ \pm$ 0.016& 0.897 $ \pm$ 0.003& 0.32 $\pm$ 0.07&0.90 $\pm$ 0.09\\ 
& F673N&&&& 0.32 $\pm$ 0.25&\\ 
180924& F160W& 0.71 $\pm$ 0.12& 3.37$ \pm$ 0.56& 1.20 $ \pm$ 0.20& 0.24 $\pm$ 0.11&0.66 $\pm$ 0.07\\ 
& F300X&&&&&\\ 
190102& F160W& 0.80 $\pm$ 0.39& 2.26$ \pm$ 2.22& 0.45 $ \pm$ 0.44& 0.39 $\pm$ 0.25&0.25 $\pm$ 0.16\\ 
& F300X&&&& 0.36 $\pm$ 0.29&\\ 
190608& F160W& 2.98 $\pm$ 0.27& 6.52$ \pm$ 0.60& 0.88 $ \pm$ 0.08& 0.19 $\pm$ 0.06&0.82 $\pm$ 0.08\\ 
& F300X&&&& 0.39 $\pm$ 0.28&\\ 
190611& GMOS-S$_r$& 2.24 $\pm$ 0.66& 11.36$ \pm$ 3.59& 5.29 $ \pm$ 1.67&&\\ 
&&&&&&\\ 
190711& F160W& 0.53 $\pm$ 0.27& 1.94$ \pm$ 2.30& 0.78 $ \pm$ 0.93& 0.55 $\pm$ 0.27&0.61 $\pm$ 0.22\\ 
& F300X&&&&&\\ 
190714& F160W& 0.61 $\pm$ 0.29& 1.97$ \pm$ 1.18& 0.51 $ \pm$ 0.31& 0.34 $\pm$ 0.23&0.23 $\pm$ 0.10\\ 
& F300X&&&& 0.38 $\pm$ 0.31&\\ 
191001& F160W& 2.74 $\pm$ 0.15& 10.49$ \pm$ 0.59& 2.87 $ \pm$ 0.16& 0.09 $\pm$ 0.03&0.91 $\pm$ 0.09\\ 
& F300X&&&& 0.29 $\pm$ 0.25&\\ 
200430& Pan-STARRS$_r$& 1.30 $\pm$ 0.79& 2.97$ \pm$ 2.36& 1.81 $ \pm$ 1.44&&\\ 
&&&&&&\\ 
\hline 
Median (IR) & & 1.1 & 3.2 & 0.9 & 0.33 & 0.6 
\\ 
16,84\% Interval & & [0.6,2.9] & [2.0,8.7] & [0.5,2.4] & [0.20,0.53] & [0.2,0.9]  
\\ 
Median (UV) & &  &  & & 0.37  
\\ 
16,84\% Interval & &  &  &  & [0.32,0.45]  
\\ 
\hline 
\enddata 
\tablecomments{FRBs\,190611 and 200430 are derived from ground-based imaging as reported in \citet{Heintz20}} 
\end{deluxetable*}


\begin{figure*}[!ht]
    \centering
    \includegraphics[width=0.49\textwidth]{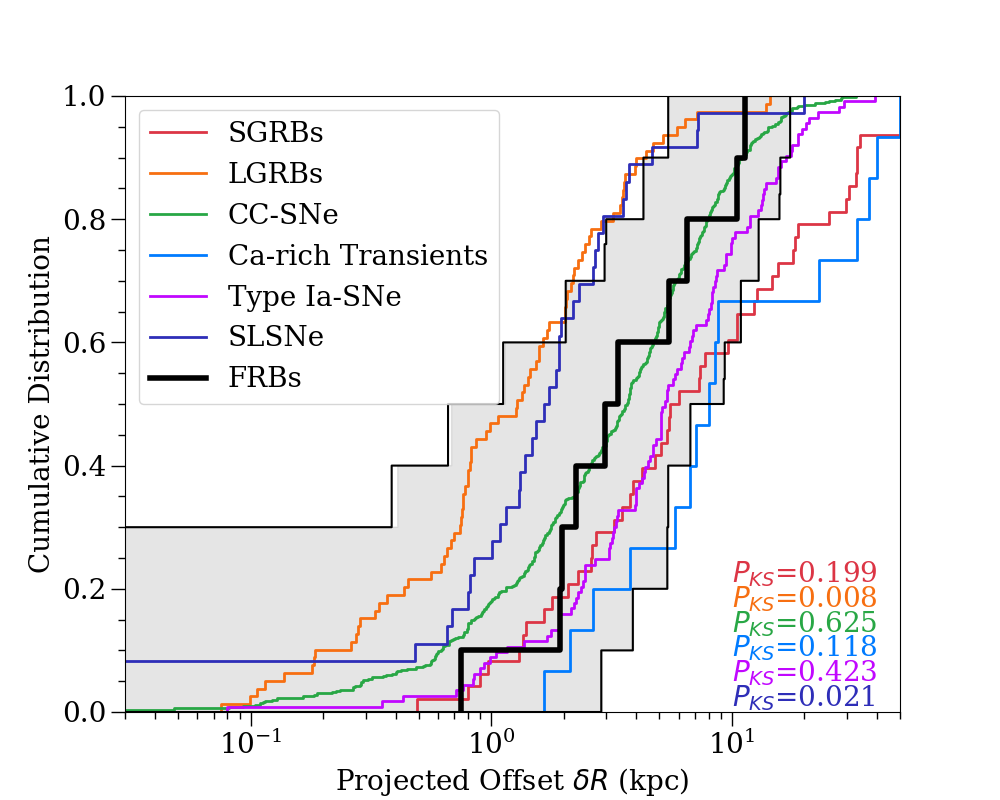}
    \includegraphics[width=0.49\textwidth]{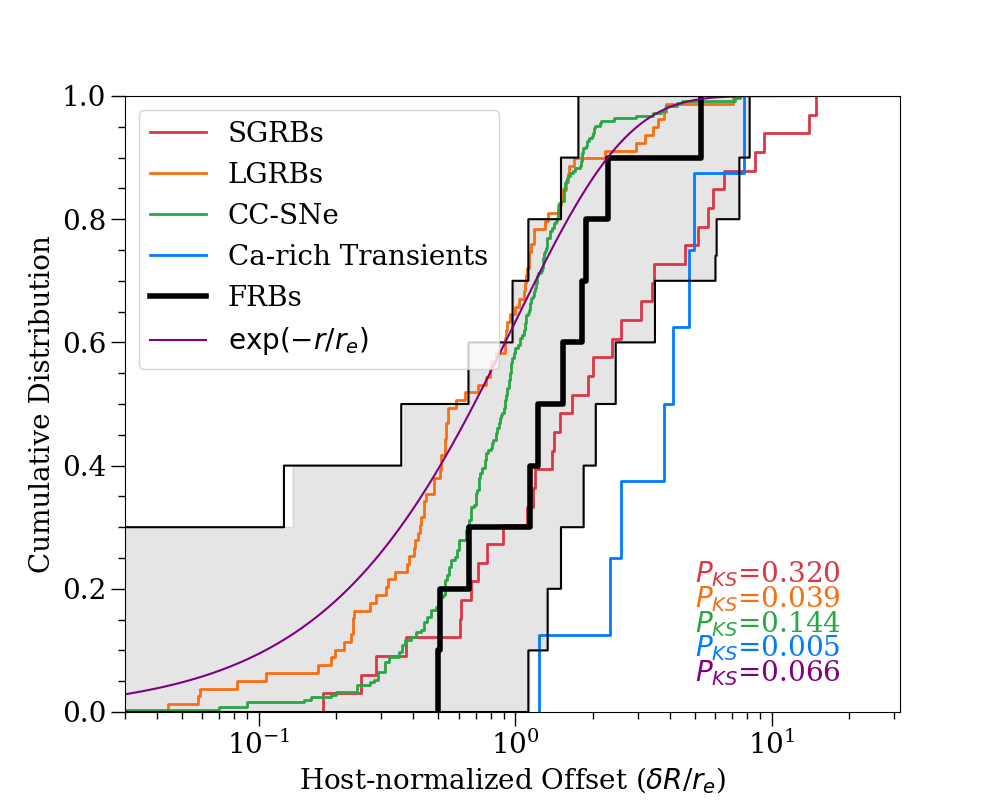}
    \caption{{\it Left:} Cumulative distribution of projected physical offsets, $\delta R$, for the 10~FRBs in the {\it HST} and ground-based samples (black line). The gray shaded region 
    is a bootstrap estimate of the RMS of the distribution,
    which accounts for both uncertainties on individual measurements, as well as statistical uncertainties due to the sample size. Comparison samples are included for SGRBs \citep{Fong10,Fong13}, LGRBs \citep{Blanchard16}, Ca-rich transients \citep{Lunnan17,De20}, Type Ia SNe \citep{Uddin_2020}, CCSNe \citep[]{Schulze20}, and SLSNe \citep{Lunnan15,Schulze20} for events at $z<1$. The computed $P$-values from a two-sided KS test are listed for each population relative to the FRB sample. {\it Right:} The same as the right-hand panel but for the host-normalized offsets ($\delta R/r_e$). This plot also shows the profile of an exponential disk.}
    \label{fig:offset_cdf}
\end{figure*}

\subsection{Physical \& Host-normalized Offsets}
\label{ssec:off}

Given that FRB localizations are typically non-circular (elliptical) in shape, and that their values span a range ($\sigma_{\rm FRB} \approx 0.0023-0.7''$), it is necessary to take their shape, size, and orientation into account when calculating the angular, physical, and host-normalized offsets.  While the synthesized beam and hence localization ellipse of the FRB can in principle have any position angle, most FRB detections have been made with beams that are close to circular and report the positional uncertainties projected onto the right ascension and declination axes, and we construct our localization ellipses using these projected values. To determine the total uncertainty on offset measurements, we sum each of the RA and Dec components of the three sources of uncertainty, $\sigma_{\rm tie}$, $\sigma_{\rm host}$, and $\sigma_{\rm FRB}$ in quadrature. We use the total uncertainties in RA and Dec to define an ellipsoidal region that represents the FRB location on the {\it HST} image. 

The estimated angular offset \angoff\ is then the convolution
of the offset from the 
galaxy centroid $\alpha_g$ with the 
FRB localization:

\begin{equation}
\mangoff = \int d\omega \, |\alpha_g - \omega|
   L(\alpha_{\rm FRB} - \omega)
\end{equation}
with $L$ a 2-D Gaussian set by the ellipsoidal region
described above.
To evaluate this convolution,
we divide each 5-$\sigma$ region around the FRB
into four million grid points 
by imposing a $2000 \times 2000$-point sub-grid. 
We measure the angular offset between each grid-point 
$i$ and the host galaxy center to obtain a distribution of angular offsets $\mangoff_i$ 
for each FRB. Finally, we apply a 2-D Gaussian probability distribution within the FRB localization ellipse, centered on the central RA and Dec of the FRB, and weight the angular offset distribution by the corresponding values. 
We estimated the variance in \angoff\ in a similar manner
and report the RMS in Table~\ref{tab:locations}.

For each FRB, we determine the median offset and standard deviation. We find a range of projected angular offsets of $\mangoff \approx 0.23-7.87''$ with a population median and 68\%\ interval in the IR of 
$\medangoff$ and $\qangoff$.
The values for each FRB are listed in Table~\ref{tab:locations}. We note that we only obtained offsets for observations in which the host galaxy center could be well determined, so this includes all hosts for which there are IR images, 
as well as the host galaxy of FRB\,190608 in both the IR and UV.

We convert the angular offsets to projected physical offsets using the redshift of each FRB host galaxy and a Planck cosmology with $H_0 = 67.8$~km~s$^{-1}$~Mpc$^{-1}$, $\Omega_M = 0.308$, $\Omega_{\Lambda}=0.692$ \citep{Planck15}. 
For the {\it HST} offsets, we find a range of $\mphysoff \approx 0.75-10.5$~kpc with the lower and upper bounds set by FRB\,121102 and FRB\,191001, respectively\footnote{We note that the ground-based determination for FRB\,190611 is the largest physical offset, with $\approx 11.4$~kpc.}. The median and 68\%\ interval on the projected physical offset 
is \medphyoff\ and \qphyoff.
Finally, we use the host galaxy half-light radii ($r_e$), as measured from {\it HST} imaging (see Section~\ref{ssec:isophote_galfit}) to determine the host-normalized offsets, $\delta R/r_e$. The values for the projected angular, physical, and host-normalized offsets for the eight FRBs in our sample are listed in Table~\ref{tab:locations}.

We supplement the FRB distributions with two FRB host galaxies in \citet{Heintz20}, FRBs\,190611 and 200430, both of which have offsets determined from ground-based imaging with seeing of $\sim\!0.8''$. To determine the uncertainty on the cumulative distribution, we follow the method by \citet{Palmerio19} and create 10,000 realizations of asymmetric Gaussian PDFs using the errors on the offset measurements for each FRB, derived from the previously described weighted grid analysis. We then use a bootstrap method to sample from the PDF in each realization, allowing us to compute a CDF of the bootstrapped sample. Finally, we compute the median of all the resulting CDFs, as well as the upper and lower bounds for each bin. We perform this same analysis for the projected physical and host-normalized offset distributions. The resulting median cumulative distributions, and bootstrap estimate of the uncertainty
(shown as the shaded gray region) are shown in Figure~\ref{fig:offset_cdf}.

To compare the FRB distribution to the offset distributions of other transients, we draw relevant comparison samples from the literature. Included are long-duration gamma-ray bursts (LGRBs; \citealt{Blanchard16}, short-duration gamma-ray bursts (SGRBs; \citealt{Fong10, Fong13}), Ca-rich transients \citep{Lunnan17,De20}, Type Ia supernovae (Type Ia SNe; \citealt{Uddin_2020}), core-collapse SNe (CCSNe; \citealt{Schulze20}) and super-luminous SNe (SLSNe; \citealt{Lunnan15,Schulze20}). To align with the redshift distributions of the FRBs, we only include values for events with $z<1$. We perform a two-sided KS-test between the median FRB distribution and each of the transient populations to test the null hypothesis that the (median) distribution of FRBs and each transient population is drawn from the same underlying distribution. Using this analysis for projected offsets, we find $P_{\rm KS}<0.05$ for both LGRBs and SLSNe, rejecting the null hypothesis that they are drawn from the same continuous distributions. We caution, however, that we only report KS-test results on the median FRB distribution.

However, for host-normalized offsets, we also find $P_{\rm KS}<0.05$ for Ca-rich transients and for LGRBs.
The remaining $p$-values are $P_{\rm KS}>0.05$, and thus we cannot reject the null hypothesis for any other population tested. Finally, we compare the distribution to an exponential disk light profile (light purple curve in Figure~\ref{fig:offset_cdf}). While the distribution overall appears to be at larger offsets, the $P_{\rm KS}=0.066$ value is not conclusive. In this analysis, we note that we are treating the FRB population as a single distribution with a dominant progenitor population. 
The sample size considered here prevents meaningful constraints on the presence of multiple, equally dominant progenitor populations, each of which have distinct offset distributions.

\subsection{Fractional Flux}
\label{ssec:ff}

\begin{figure*}[ht]
    \centering
    \includegraphics[width=0.49\textwidth]{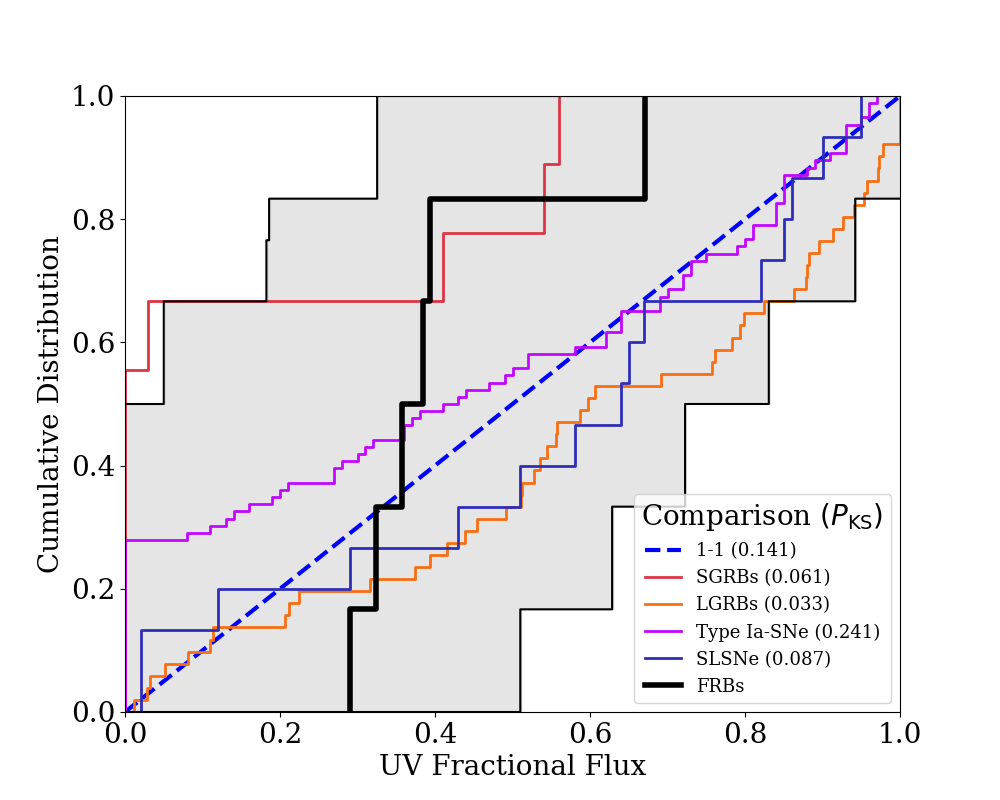}
    \includegraphics[width=0.49\textwidth]{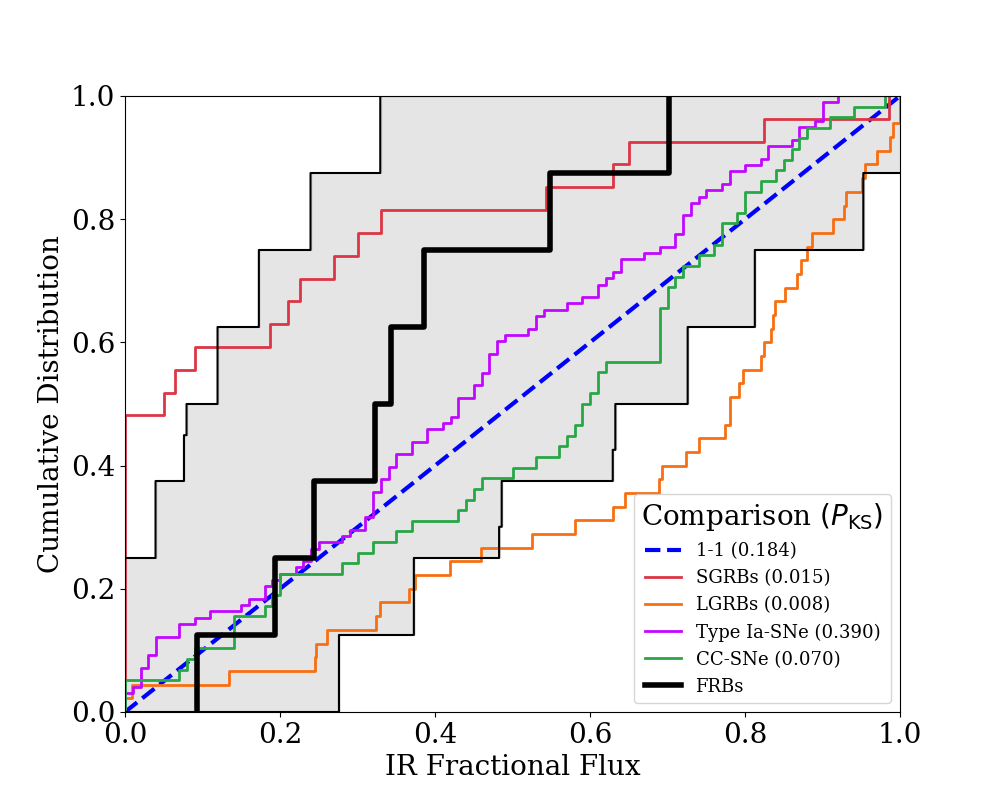}
    \caption{{\it Left:} Cumulative distribution of UV fractional flux (\FF) of the 6 FRB hosts in the {\it HST} sample with WFC3/UVIS imaging. The gray shaded region is a bootstrap estimate of the RMS of the distribution,
    which accounts for both uncertainties on individual measurements, as well as statistical uncertainties due to the sample size. For comparison, the corresponding distributions for SGRBs \citep{Fong10,Fong13}, LGRBs \citep{Blanchard16}, Type Ia-SNe \citep{Wang13}, CCSNe \citep{Svensson10}, and SLSNe \citep{Lunnan15} are shown. Also shown is the dashed, 1:1 line, representing the distribution of host galaxy light. The $p$-values from two-sided KS tests between each population and the median FRB distribution are listed. {\it Right:} The same as the left-hand panel but for all eight FRB hosts with {\it HST}/IR imaging.}
    \label{fig:ff_cdf}
\end{figure*}

We now explore the location of the FRBs with respect to their host galaxy light distributions (``fractional flux''; \FF). The brightness of the burst site in relation to how its rest-frame UV and optical host light is distributed is a crucial tool for determining how star formation activity and stellar mass are tracked \citep{Fruchter06}. Compared to offsets, which can depend on host size and morphology, the fractional flux method is independent of these physical characteristics. Specifically, the measurement determines the fraction of host light fainter than the flux at the burst position, where a value of unity corresponds to the brightest light level of the host ($\mFF = 1)$.

Foreground stars in the field of FRB\,180916 preclude the direct application of this analysis to the field. Since PSF photometry and subtraction of many of these stars were difficult as they were saturated; in this case we decided to use an alternate, ``brute-force'' approach. We first performed an isophote fit to the galaxy using \texttt{photutils} \citep{photutils}, clipping pixels which were over the $3\sigma$ level compared to the local mean. This left all the foreground stars and subtracted most of the galaxy light. Then we created a segmentation map from the residual image with a threshold level of $4\sigma$ over the sky background and minimum source area of 5 pixels (the default value). From the objects extracted, we selected those with a peak pixel value of 1~e$^{-}$ per sec or greater and created a masked image. This masked out all star light above the segmentation map threshold value. We then replaced the masked pixels with the isophote fit from earlier and used this stitched image for the fractional flux analysis.

We center a 2D cutout on each host galaxy, making sure that we include a sufficient amount of background. We then determine which pixels lie within the 3-$\sigma$ FRB localization ellipse. We note that unlike how we determined the offsets (Section~\ref{ssec:off}), we do not apply a sub-grid to the localization, as the main limitation is the pixel scale, and we cannot resolve the fractional flux below this scale. The fractional flux for each $i$th pixel on N pixels within the localization is then calculated as 

\begin{equation}
    \mFF =\frac{\sum\limits_i(F_{i} < {\rm limit})}{\sum\limits_i F_{i}}.
\end{equation}

\noindent We then use a 2-D Gaussian distribution to develop a weighting scheme with each FRB localization ellipse in the same manner as that described in Section~\ref{ssec:off}. From the distribution of \FF\ values for each FRB, we determine the median \FF\ and its standard deviation. 
The values for the fractional flux for each FRB can be found in Table~\ref{tab:locations}.

Figure~\ref{fig:ff_cdf} shows the \FF\ CDF for the sample of eight {\it HST} hosts from the IR images (corresponding to rest-frame optical; right panel), and four hosts from the UV images (left panel). 
The gray region was generated in the same manner described in Section~\ref{ssec:off}.
We exclude the UV imaging of FRBs\,180924 and 190711 from this analysis because they are effectively non-detections with very large error, while we include the $H\alpha$ imaging for FRB\,180916 as a proxy for star formation \citep{tendulkar20}. 
Given that the UV and IR imaging can be used as proxies for star formation and stellar mass, respectively, we keep the two wavelength regimes separated. 
In this method, the 1:1 dashed line represents a population of events which traces the light of its host galaxy in that band. Adherence to the 1:1 line would indicate that FRBs may trace the distribution of star-forming regions and stellar mass of their galaxies, respectively.

For the IR (rest-frame optical) distribution which can be used as a proxy for stellar mass, the \FF\ values span a wide range, $\approx 0.09-0.70$, where the lower and upper bounds are set by FRB\,190714 and FRB\,121102, respectively (Table~\ref{tab:locations}). The median of the distribution is \medFF\ with a $68\%$ interval of \qFF.

Figure~\ref{fig:ff_cdf} highlights that the median distribution of FRBs overall traces the fainter rest-frame optical regions of their host galaxies, with a location to the left/above of the 1:1 line. However, the relatively large positional uncertainties which extend to $\sigma_{\rm FRB} \approx 0.7''$, coupled with the small sample size of eight events leads to a non-trivial uncertainty in the distribution, which is consistent with the 1:1 line. Thus, while FRBs appear to trace the fainter regions of their hosts in terms of stellar mass, at present it is not possible to make a strong statistical statement. 

For comparison, we draw \FF\ measurements from the literature for SGRBs \citep{Fong10,Fong13}, LGRBs \citep{Blanchard16}, SLSNe \citep{Lunnan15}, CCSNe \citep{Svensson10}, and Type Ia SNe \citep{Wang13} and divide them into rest-frame UV and optical measurements for direct comparisons. Performing two-sided KS tests with respect to the median FRB distribution, we can rule out the null hypothesis that the LGRBs, SGRBs and FRBs are from the same underlying population as they all yield $p$-values of $P_{KS} < 0.05$. For CCSNe, Type Ia SNe, and the 1:1 distribution, we find $P_{KS}>0.05$ and cannot rule out the null hypothesis. 

For the UV distribution, which can be used as a proxy for the distribution of current star formation, there are six data points and their median values fall in a fairly narrow range of $\approx 0.29$ (for FRB\,191001) to $\approx 0.67$ (for FRB\,121102). Despite the small sample size, the KS test does reject the null hypothesis that both LGRBs and SLSNe and the median FRB distribution come from the same underlying population with $P_{KS}<0.05$. However, we caution that the very small sample size coupled with large localization uncertainties (which translate to large measurement uncertainties in \FF) effectively means that almost the entire parameter space of \FF\ is included in the uncertainty region. We also note that the IR \FF\ contains all three repeating FRBs, whereas the UV \FF\ contains two.

\subsection{Fraction of enclosed light}
\label{ssec:f_enclosed}

\begin{figure}[!t]
    \centering
    \includegraphics[width=0.48\textwidth]{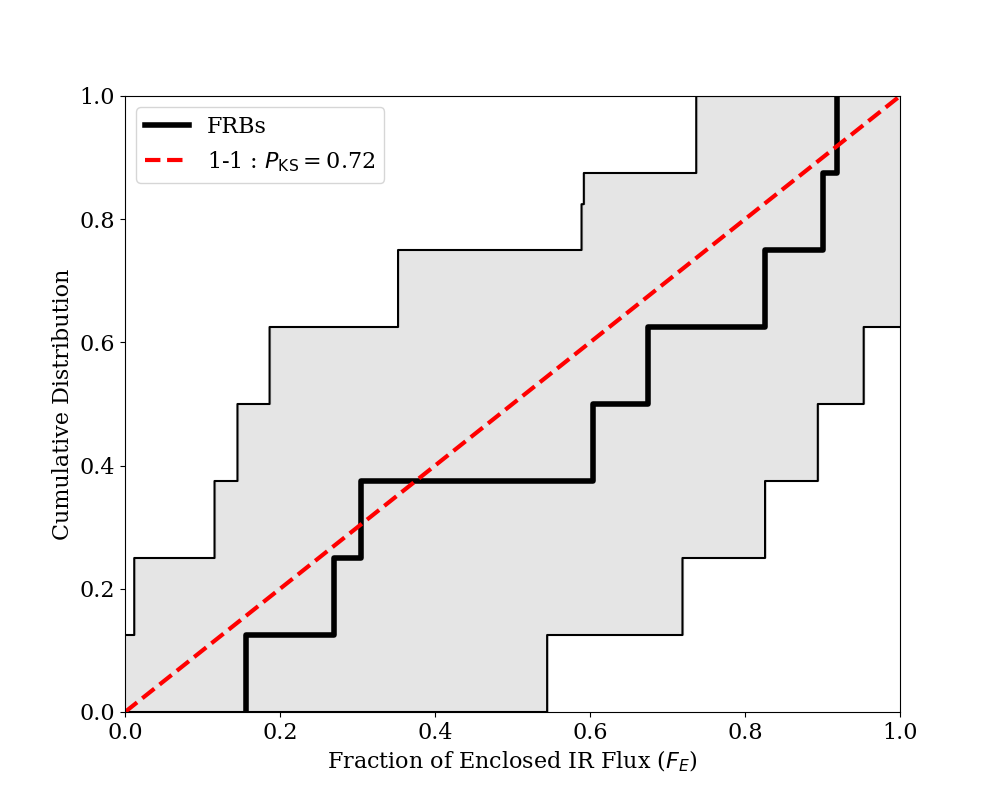}
    \vspace{-0.2in}
    \caption{The median FRB cumulative distribution of enclosed flux (black line), which is the fraction of host light enclosed within a radius set by the position of the FRB. The values are derived for eight host galaxies with IR images. The gray shaded region represents the $1\sigma$ uncertainty in the CDF, while the blue dashed line indicates a 1:1 mapping. A KS test between the median distribution and 1:1 line does not reject the null hypothesis that the distributions are drawn from the same underlying population.
    }
    \label{fig:f_enclosed}
\end{figure}

While the \FF\ metric is designed to assess the local environment
of the FRB in the galaxy {\it independent} of its morphology,
we introduce an alternative metric to better assess its
global position: the fractional flux enclosed

\begin{equation}
    \mFE = \frac{\sum\limits_{r_i < r_{\rm FRB}} F_i}{
    \sum\limits_i F_i}
\end{equation}
with $r_i$ the radius of pixel~i and
$r_{\rm FRB}$ the distance of the FRB from the
galaxy centroid.
In practice, we estimate \FE\ using the isophotal analysis
of Section~\ref{ssec:isophote_galfit}.
Specifically, we find the isophote closest to the FRB
localization and measure the flux enclosed within it
(and all interior isophotes).  Furthermore, we allow for
the FRB localization error by performing a 
weighted average of individual \FE\ evaluations
across the FRB localization.

Figure~\ref{fig:f_enclosed} shows the \FE\ results
derived from the WFC3/IR images and a 1:1 line corresponding
to the null hypothesis that FRBs are biased in tracking the host's light.
The close correspondence between the two lends credence
to the null hypothesis and the resultant 
$P_{\rm KS} = 0.73$ value offers statistical support.
Similar to \cite{Safarzadeh20}, we infer that FRBs 
track the general distribution of light and, by 
inference, stellar mass in their host galaxies.

\section{Morphological \& FRB Site Properties}
\label{sec:explosion_sites}

\begin{deluxetable*}{cccccccccccccccc}
\tablewidth{0pc}
\tablecaption{Derived Properties from UV Observations of Host Galaxies\label{tab:UVproperties}}
\tabletypesize{\normalsize}
\tablehead{\colhead{FRB}  
& \colhead{Filter} 
& \colhead{Host Magnitude}  
& \colhead{$A_\lambda$}  
& \colhead{$\mu_{\rm FRB}$}  
& \colhead{$\Sigma_{\rm SFR(FRB)}$}  
\\&  &  (AB mag) & (mag) & ($\mu$Jy arcsec$^{-2}$) 
& ($M_\odot \rm \, yr^{-1} \, kpc^{-2})$ 
} 
\startdata 
180924& F300X& 23.478 $\pm$ 0.058& 0.12& $<$ 0.85 & $<$ 0.006\\ 
190102& F300X& $>$ 27.200& 1.40& $<$ 2.58 & $<$ 0.016\\ 
190608& F300X& 19.765 $\pm$ 0.014& 0.30& 1.73 $\pm$ 0.033 & 0.007 $\pm$ 0.001\\ 
190711& F300X& 25.008 $\pm$ 0.121& 0.88& $<$ 1.61 & $<$ 0.016\\ 
190714& F300X& 23.072 $\pm$ 0.053& 0.39& $<$ 1.12 & $<$ 0.006\\ 
191001& F300X& 21.228 $\pm$ 0.020& 0.18& $<$ 0.88 & $<$ 0.005\\ 
\hline 
\enddata 
\tablecomments{Magnitudes are not corrected for Galactic extinction in the direction of the FRB 
($A_\lambda$). Limits correspond to $5\sigma$ confidence.} 
 \end{deluxetable*} 

\begin{deluxetable*}{cccccccccccccccc}
\tablewidth{0pc}
\tablecaption{Derived Properties from IR Observations of Host Galaxies\label{tab:IRproperties}}
\tabletypesize{\normalsize}
\tablehead{\colhead{FRB}  
& \colhead{Filter} 
& \colhead{$r_e$ (Isophotal)}  
& \colhead{$r_e$ ({\tt GALFIT})}  
& \colhead{Host Magnitude}  
& \colhead{Limit}  
& \colhead{$A_\lambda$}  
& \colhead{$\mu_{\rm FRB}$}  
& \colhead{$\Sigma_{M*(FRB)}$}  
\\ 
&  &  (kpc) & (kpc) & (AB mag) & (AB mag) & (mag) & ($\mu$Jy arcsec$^{-2}$) 
& ($10^8 \, M_\odot \rm \, kpc^{-2})$ 
} 
\startdata 
121102& F160W& 0.99 $\pm$ 0.07& 2.05 $\pm$ 0.11& 23.435 $\pm$ 0.055& 27.4& 0.40& 14.57 $\pm$ 0.20& 0.130 $\pm$ 0.002\\ 
180916& F110W& 2.52 $\pm$ 0.15& 6.00 $\pm$ 0.01& 16.178 $\pm$ 0.005& 25.4& 0.88& 31.00 $\pm$ 0.10& 0.115 $\pm$ 0.000\\ 
180924& F160W& 2.54 $\pm$ 0.41& 2.82 $\pm$ 0.53& 19.349 $\pm$ 0.002& 26.2& 0.01& 20.69 $\pm$ 0.10& 0.810 $\pm$ 0.010\\ 
190102& F160W& 4.87 $\pm$ 0.34& 5.00 $\pm$ 0.15& 20.550 $\pm$ 0.006& 27.1& 0.11& 11.10 $\pm$ 0.11& 0.093 $\pm$ 0.002\\ 
190608& F160W& 2.65 $\pm$ 0.53& 7.37 $\pm$ 0.06& 16.693 $\pm$ 0.001& 25.2& 0.02& 17.80 $\pm$ 0.10& 0.340 $\pm$ 0.001\\ 
190711& F160W& 2.88 $\pm$ 0.32& 2.48 $\pm$ 0.13& 22.899 $\pm$ 0.014& 27.6& 0.07& 6.17 $\pm$ 0.11& 0.045 $\pm$ 0.004\\ 
190714& F160W& 4.07 $\pm$ 0.35& 3.85 $\pm$ 0.03& 18.896 $\pm$ 0.002& 25.9& 0.03& 18.00 $\pm$ 0.10& 1.752 $\pm$ 0.018\\ 
191001& F160W& 6.05 $\pm$ 0.66& 6.23 $\pm$ 0.04& 17.135 $\pm$ 0.001& 24.8& 0.01& 7.99 $\pm$ 0.10& 0.322 $\pm$ 0.007\\ 
\hline 
\enddata 
\tablecomments{Magnitudes are not corrected for Galactic extinction in the direction of the FRB 
($A_\lambda$). Limits correspond to $5\sigma$ confidence.} 
 \end{deluxetable*} 



\subsection{Galaxy Light Profile Fitting} \label{ssec:isophote_galfit}
We fit the light profile for the eight FRB hosts with {\it HST} data to determine the half-light radii ($r_e$), which are used to compute the host-normalized offsets. The half-light radii from {\it HST} imaging are valuable in comparison to previous ground-based imaging, where the measurements are based on seeing-limited images, allowing for more accurate estimates. The increased sensitivity of {\it HST} also presents the opportunity to search for alternate, fainter host galaxy candidates. Thus, in what follows, we also use light profile fitting to develop a galaxy model, and subtract it from the images to determine constraints on possible alternative, host galaxy candidates (e.g., low-luminosity galaxies or background galaxies) at the position of each FRB. 

First, we use the elliptical isophotal model from $\texttt{photutils.isophote}$ to map the light of the eight FRB hosts. We begin with an initial guess, providing values for the central position, ellipticity, semi-major axis, and position angle. The function then fits a series of isophotes which we then use to create a model and residual image. We determine the value of $r_e$ from our isophotal fits, taking this to be the semi-major axis in which half of the total light is enclosed. These values are listed in 
Tables~\ref{tab:UVproperties}-\ref{tab:IRproperties},
and are used in our calculation of the host-normalized offsets (see Section~\ref{ssec:off}).

In addition to the isophotal fits, we also compute residuals from S\'{e}rsic profile fits using \galfit~v.3 \citep{GALFIT} for the IR images. Our model has two S\'{e}rsic components for all galaxies except the hosts of FRBs\,180924 and 190711. These two components roughly correspond to a central bulge and an outer disk. In the case of FRB\,180924, we use a single S\'{e}rsic component 
because there is no obvious improvement in the residuals relative to performing a multi-component fit. In the case of the host galaxy of FRB\,180924, the two components converged to the same effective radii, similar S\'{e}rsic indices and similar magnitudes, implying that \galfit~could not distinctly identify separate core and disk components.
For FRB\,190711, the relatively low signal-to-noise of the host galaxy image precludes the identification of two distinct components. 

\begin{figure*}[t]
    \centering
    \includegraphics[width=\textwidth,trim={0.1in 0.2in 0.2in 0.8in},clip]{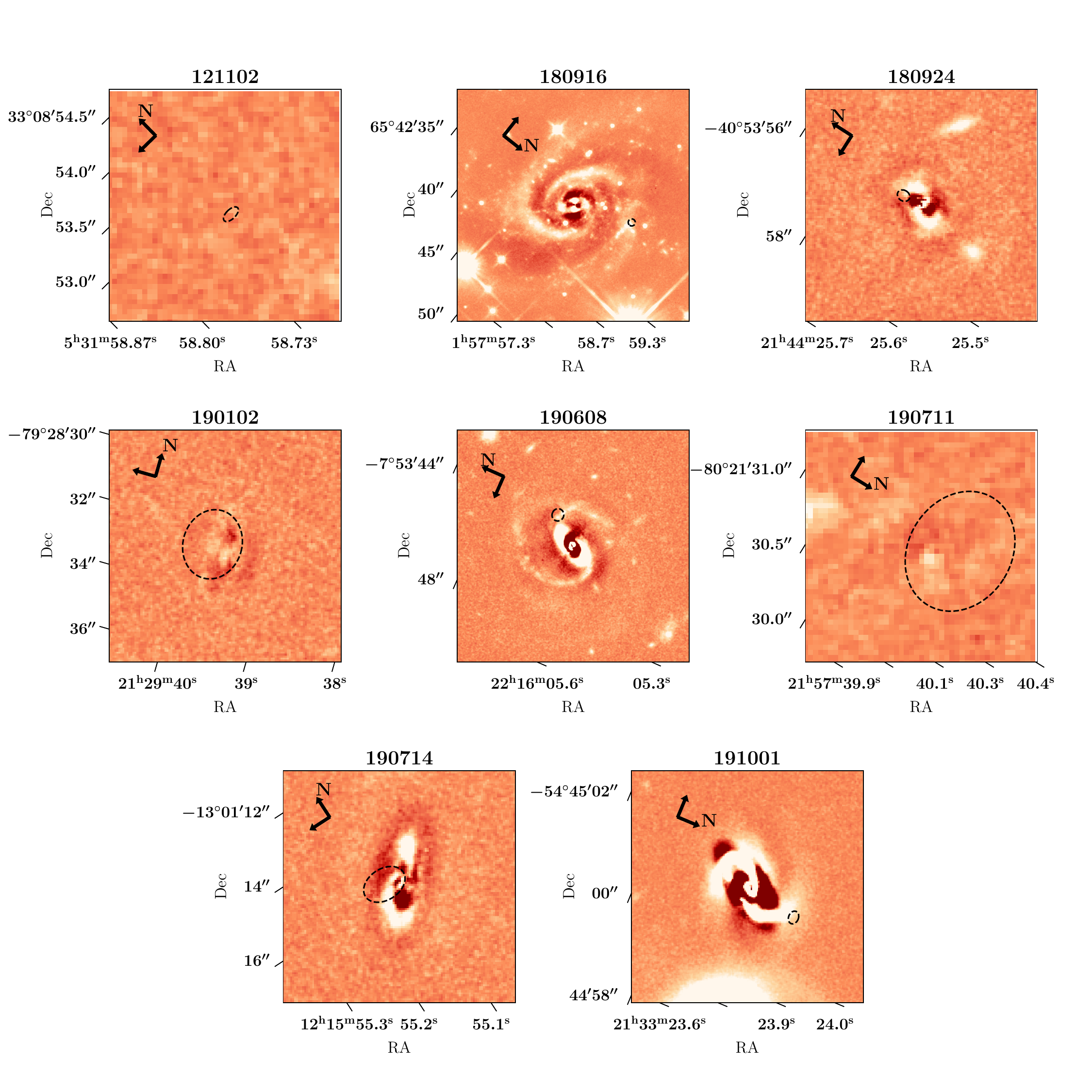}
    \vspace{-0.2in}
    \caption{Residual images produced by \galfit~from the F160W host galaxy image set (and FRB\,180916 for F110W). The North and East directions are indicated by the black arrows at the top left.  The light distribution for all galaxies was modeled as a sum of two S\'{e}rsic profiles corresponding to a central core and an outer disk, except in the host galaxies of FRBs\,180924 and 190711. In those two galaxies, a single S\'{e}rsic profile fit was used because a two component fit did not produce a significant improvement in the visual quality of the residuals.  It is interesting to note that five of the eight FRB locations, marked by the dashed 2$\sigma$ localization ellipses (200$\sigma$ for 180916 and 30$\sigma$ for 121102), are coincident with spiral structures in their respective hosts.}
    \label{fig:galfit_residuals}
\end{figure*}

The residual images from the fits are shown in Figure~\ref{fig:galfit_residuals}. The \galfit~half-light radii ($r_e$) reported in Table \ref{tab:IRproperties} correspond to the larger component (i.e. the disk component) in the case of the two-component fits.  While the residual images of FRBs\,121102, 190711 and 190102 do not show clear, symmetric structures, there are such morphological structures in the residual images of the remaining hosts. In particular, spiral arm structure becomes very apparent for FRB\,180924 and 190714, as well as the previously-known structure for FRBs\,180916 \citep{tendulkar20}, 190608 \citep{Chittidi20} and 191001 \citep{Bhandari20b}. Therefore we can see in Figure~\ref{fig:galfit_residuals} that \textit{all} of the FRBs which are localized to hosts with spiral structure land on or very near to a spiral arm; this point is discussed further in Section~\ref{sec:discuss}.

\subsection{Star Formation Rate and Stellar Mass Constraints}
\label{ssec:photo}

\begin{figure*}[!t]
    \centering
    \includegraphics[width=\columnwidth]{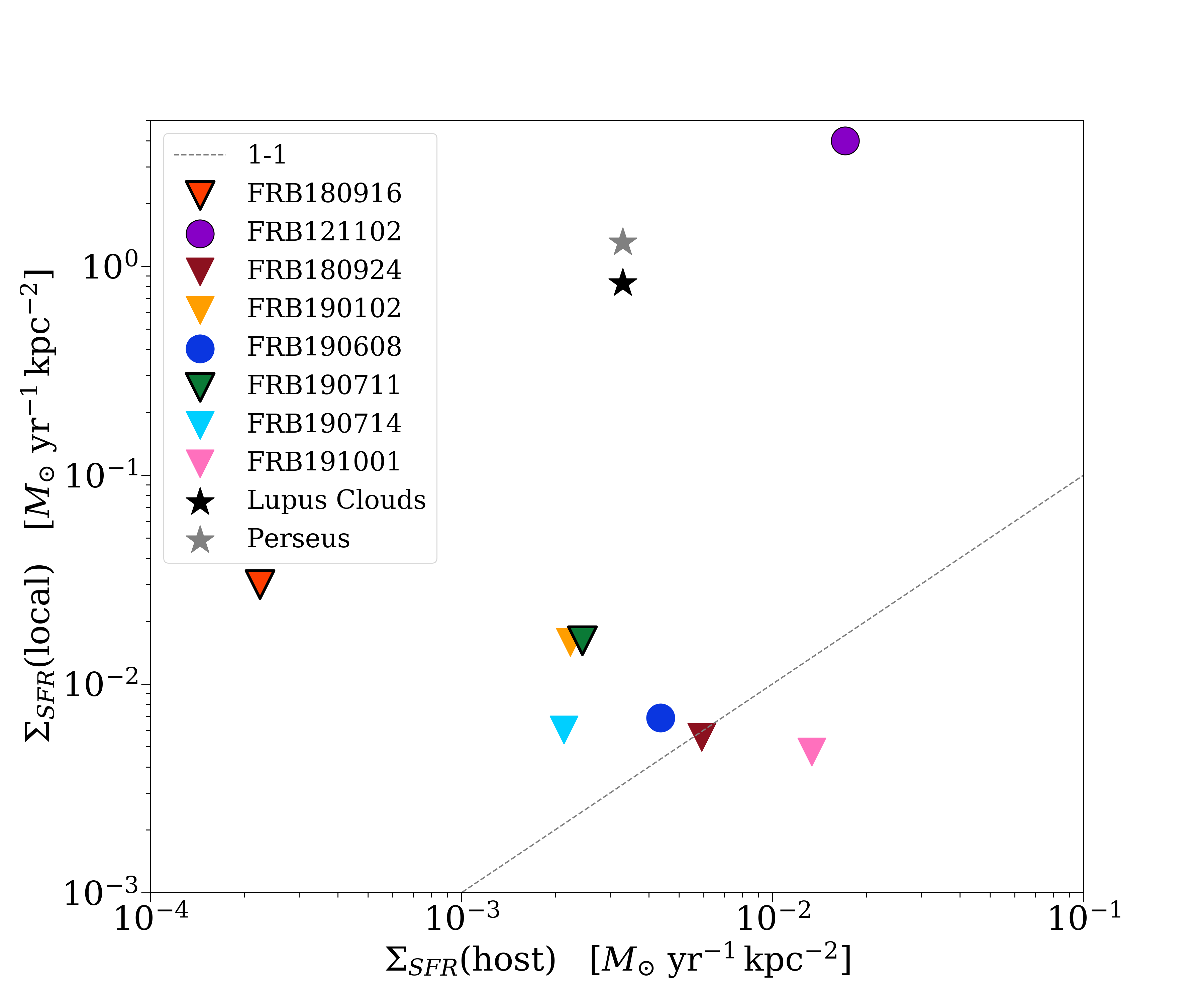}
    \includegraphics[width=\columnwidth]{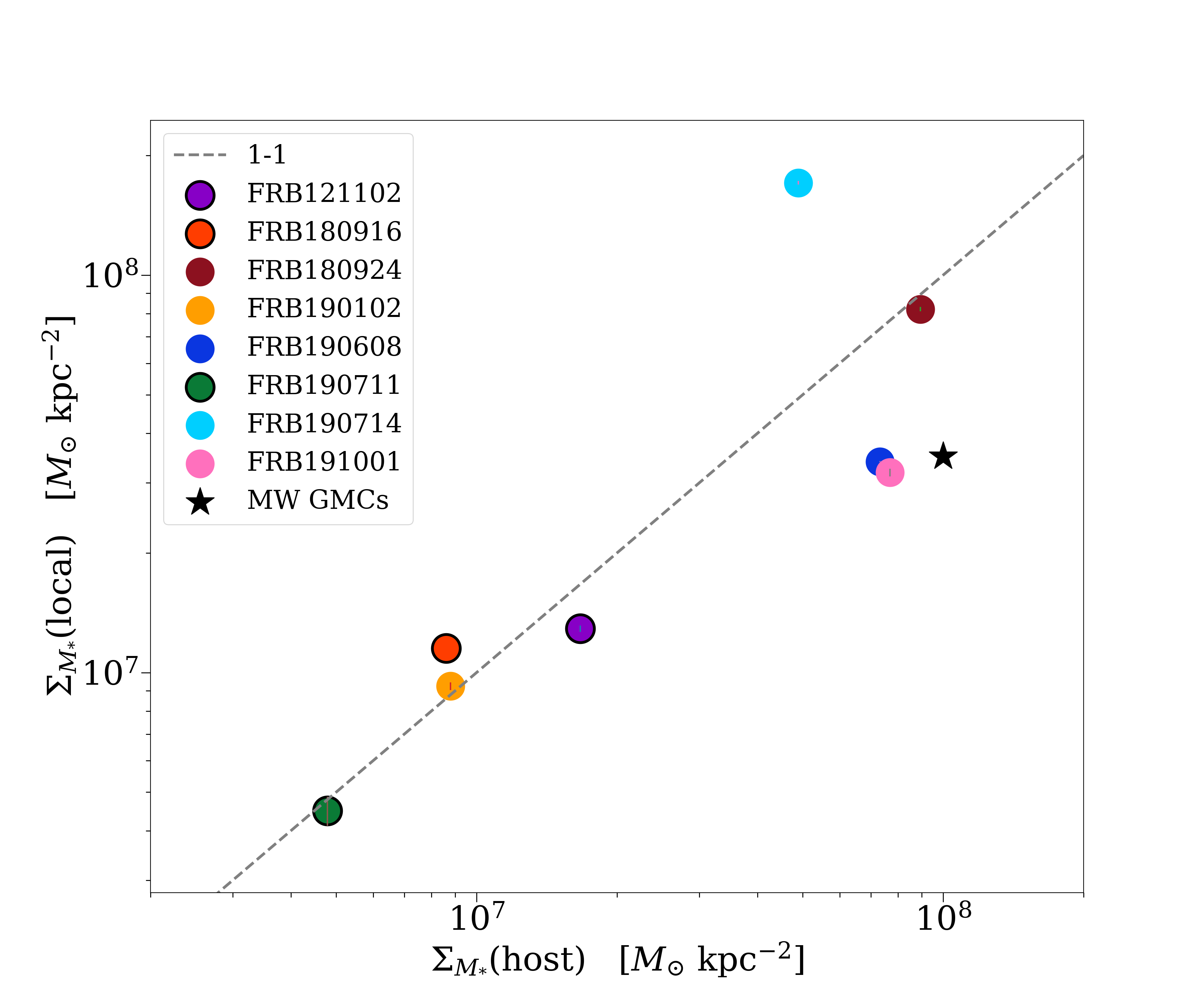}
    \caption{Comparisons of local FRB properties to global host properties. The points with black outlines are ``repeaters''. The triangles denote 3$\sigma$ upper limits on the SFR. \textit{Left:} Star formation rate surface density at FRB location versus star formation rate surface density of the host. Compared to the 1:1 line, the sites of FRBs are not clearly elevated in star formation rate surface densities with respect to their hosts, and do not reach the surface densities of Galactic star-forming regions \citep{Evans09}. \textit{Right:} Average stellar mass surface density at the burst site versus average stellar mass surface density of the host. Most FRB locations track the 1:1 line, with a few deviating from this relation. The average value for Milky Way GMCs is marked shown as the black star \citep{Lada20}.}
    \label{fig:SFR_compare}
\end{figure*}

In addition to relative photometric measures,
these data enable precise
photometric measurements at the burst positions. 
Furthermore, we may convert these light measurements
into physical quantities -- 
the star-formation rate (SFR) density \ssfr\ 
and stellar mass surface density \msd\ - to give 
additional insight into the local properties of FRB burst sites. It is also informative to compare the local values to the global mean surface densities of their host galaxies, to understand if the locations of FRBs are, for example, elevated or not in terms of these quantities.

To complete the photometric measurements, we create apertures 
with $r$ = 3 pixels at each pixel within the FRB localization. These aperture sums are then weighted by a 2-D Gaussian probability distribution centered on the measured FRB position, the same as that used for the offset and fractional flux determinations in Sections~\ref{ssec:off}-\ref{ssec:ff} -- with the resolution of the grid being limited by the image pixel scale. We then take the weighted average and divide by the area of the aperture to get an aperture sum per arcsec$^{2}$. We compute the magnitude using WFC3 tabulated zeropoints, the corresponding flux in both filters, and the luminosity for the UV band (Table~\ref{tab:UVproperties}). 

Next, we use the UV luminosity-SFR relation from \citet{Kennicutt98} to convert UV measures
into star formation rate and subsequently the star formation rate surface density per kpc$^{2}$,  
\ssfr\ at the burst site in  6 of the HST hosts. For FRB\,121102, we use the H-alpha luminosity-SFR relation from \citet{Kennicutt98b}, to obtain a value of 3.99 M$_{\odot}$ yr$^{-1}$ kpc$^{-2}$.  For the stellar mass surface density per kpc$^{2}$,  
\msd\ we compute the ratio 
of FRB flux to total host flux 
and adopt the total stellar mass estimates
from \citet{Heintz20} to estimate the local value. We derive $3\sigma$ upper limits on star formation rate densities in the same manner, relevant for hosts that are non-detections or have very low S/N at the FRB site.

In terms of \ssfr\, Figure~\ref{fig:SFR_compare} shows that most FRBs do not obviously occur in elevated regions of star formation with respect to the global values of their host galaxies (albeit most of the UV values are upper limits). The two FRBs with measurements in our sample, FRBs\,121102 and 190608 lie above the 1:1 line, in concert with previous conclusions \citep{Bassa17,Chittidi20}. We do however find that FRB\,191001 lies below this 1:1 line. This burst also has the highest offset and, as is apparent in the UV image, is offset from the UV-bright regions of its host.

For context, Galactic star-forming clouds such as the Lupus and Perseus clouds \citep{Evans09} are shown to be well above the 1:1 line in comparison to the Milky Way average as reported in \citet{Kennicutt12}. The FRB locations, except that of FRB\,121102, do not reach these levels of elevated star-formation.

One caveat is that the Galactic star forming clouds and their measurements are derived from pc-scale measurements as opposed to the kpc scales for FRB localizations. Ideally, we would like resolve down to scales that are considered ``local'' in studies of star formation in the Milky Way.
This will require additional $\sim$mas-level localizations and
larger aperture space or ground-based observations
(e.g.\ JWST, ALMA).

In terms of stellar mass surface density \msd\, Figure~\ref{fig:SFR_compare} 
reinforces several of the conclusions from \citet{Heintz20}. 
For example, a continuum of characteristics between ``repeaters'' and ``non-repeaters'' arises when investigating the stellar mass of the host and the burst site. Like the SFR density, FRBs also do not clearly occur in regions of elevated stellar mass surface densities with respect to the global values of their hosts, and only a few FRBs deviate from the 1:1 relation. We use Milky Way Giant Molecular Clouds (GMCs; \citealt{Lada20}) as a point of comparison to put into context the characteristics of these burst sites with other sites of star-formation. The stellar mass surface density for these sites relative to their hosts is slightly above, but are not very disparate from that shown for the Galactic GMCs which are $\approx35~M_{\odot}$ parsec$^{-2}$ as concluded in \citet{Lada20}. 

\subsection{Luminosity Constraints on Satellite or Background Galaxies}

{With the early association of FRB~121102 to a very faint host,
the community was led to expect 
that other FRBs would be found in galaxies of similar type.  The subsequent association
of FRBs to brighter galaxies \citep[e.g.][]{Bannister19}
has therefore led some to question whether a fainter,
true host galaxy lurks below.
To place constraints on an alternate, apparently fainter host galaxy candidate at the FRB position, we use the \galfit~residual images
(Figure~\ref{fig:galfit_residuals}), 
in which the elliptical components from the bright, putative host galaxy have been removed to derive point-source limiting 
magnitudes \mlim\ at the FRB position. We then compute the residual flux value using a circular aperture of $0.5''$ diameter, corresponding to $\sim 2.5$ times the PSF FWHM. 
We compute the net standard deviation for all pixels within this aperture. We then take the larger of the flux measurement and five times the net standard deviation as the upper limit on any point source flux that can be detected from the residual images ($5\sigma$ limit). We find limits of $\mmlim \gtrsim 24.8-27.6$~AB\,mag 
(see Table \ref{tab:IRproperties}).

We convert each of the at-position limits to an IR luminosity as a function of redshift (Figure~\ref{fig:lumz}). First, we explore these limits in the context of a spatially coincident satellite galaxy at the same redshift as the putative (brighter) host galaxy (triangles in Figure~\ref{fig:lumz}). At these redshifts, the limits of $L_{\rm IR} \lesssim (0.5-9.2) \times 10^{7}~L_{\odot}$, are significantly deeper than the luminosity of any known FRB host, including FRB\,121102.  This means that despite the presence of morphological features in the \galfit~residuals which preclude extremely deep limits, we can still rule out a galaxy with similar luminosity to the host of FRB\,121102, which is now considered an outlier in terms of FRB host stellar mass and luminosity \citep{Tendulkar17,Li19,Bhandari20a,Heintz20}. Any underlying host would need to have an IR luminosity of $\lesssim 0.02-0.31$~times that of the host of FRB\,121102 if it was at the same redshift of the brighter host galaxy (Figure~\ref{fig:lumz}). For reference, we measure IR luminosities for the putative hosts of $\approx 3.0 \times 10^{8} - 1.1 \times 10^{11}~L_{\odot}$ (set by FRB\,121102 and FRB\,191001, respectively). 

It is also worthwhile to explore whether or not a low-luminosity host galaxy of the same luminosity as the host of FRB\,121102 may reside at a {\it higher} redshift than the apparently brighter galaxy. In this case, we find that the redshift of any background galaxy must be at $z \gtrsim 0.4$ (Figure~\ref{fig:lumz}). The exception is FRB\,180916, which still has a meaningful constraint of $z \gtrsim 0.25$. We also calculate the upper limit on the redshift inferred from the Macquart relation for each FRB following the methods of \citet{Macquart20}, and assuming a Milky Way DM = 50~pc~cm$^{-3}$ and a host DM = $50/(1+z_{\rm FRB})$~pc~cm$^{-3}$. This analysis results in limits of $z \lesssim 0.17-0.75$ (95\% confidence). These redshift limits provide an absolute upper bound on the allowed luminosity of an underlying host of $\approx 4 \times 10^{7} - 8 \times 10^{8}\,L_{\odot}$. In all cases except for FRBs\,190714 and 191001, we can thus rule out an underlying background galaxy of similar luminosity to the host of FRB\,121102. We note that raising the required luminosity of the host galaxy would only push the required redshift to a higher range. We therefore find that the presence of background galaxies at higher redshifts are not likely for these FRBs given the {\it HST} limits and constraints from the measured DMs of the FRBs.

\begin{figure}[t]
    \centering
    \includegraphics[width=0.5\textwidth]{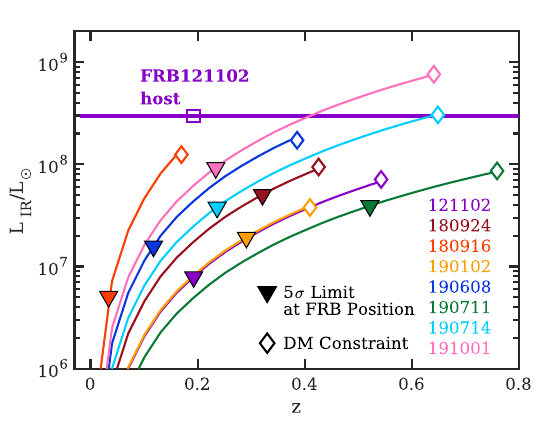}
    \vspace{-0.3in}
    \caption{Limits on the near-IR luminosity at the FRB positions (lines) as a function of redshift, derived from \galfit~residual images after a S\'{e}rsic component is removed. Filled triangles represent $5\sigma$ limits at the redshift of the putative, brighter FRB host galaxy. The limits rule out satellite galaxies at the FRB positions at the same redshift as the bright host to deep limits. If instead a background galaxy exists with a host luminosity similar to FRB\,121102, this would require redshifts larger than allowed by the measured DM (diamonds, 95\% confidence) in all cases except FRB\,190714 and 191001.}
    \label{fig:lumz}
\end{figure}


\section{Discussion}
\label{sec:discuss}

Here we discuss the locations, luminosity limits, and morphological features revealed by {\it HST} imaging in the context of other transient populations with known progenitors, and implications for FRB progenitors.

\subsection{The Locations of FRBs with Respect to their Host Galaxies}

The high angular resolution of the {\it HST} imaging enables the determination of effective radii and the precise locations for FRB events with respect to their host galaxies using a variety of measures. In general, locations have been used in a variety of transient studies as a major diagnostic to uncovering their progenitors \citep{Prieto08,Fong10,Fong13,Blanchard16,Lunnan17,De20,Audcent-Ross20,Schulze20}, as well as the relation to the distribution of young stars (UV, H$\alpha$ light) and stellar mass (IR light) in their host galaxies. As a means to deciphering the origins of FRBs, we consider comparison data sets from transients which span a wide range of progenitor systems, from those which originate from massive stars in which the populations follow the UV light and exponential disk profiles of their hosts (LGRBs, CCSNe, SLSNe; \citealt{Fruchter06,Lunnan15}); to those with older stellar progenitors associated with compact objects which are weakly correlated with the UV light of their hosts (SGRBs, Type Ia SNe; \citealt{Fong13,Wang13,Audcent-Ross20}); to those with unknown progenitors and larger offsets from their hosts (Ca-rich transients\footnote{These are also referred to as Ca-strong transients or CASTs; \citet{Shen19}}).

Comparative studies based on ground-based observations of FRBs have found that the spatial distribution of FRBs are inconsistent with the distribution of LGRBs and SLSNe, both of which originate from stripped-envelope massive stars \citep{Fruchter06,Lunnan15}, but are consistent with other transient types \citep{Bhandari20a,Heintz20, Bochenek20}. 
Our results support these studies, where we find that the locations of FRBs as a population are clearly more extended than LGRBs and SLSNe in terms of physical offsets, with a median of \medphyoff\ (68\% interval of \qphyoff). However, the host galaxies of FRBs are on average larger in physical size (and also stellar mass; c.f., \citealt{Bhandari20a,Heintz20}) than the hosts of LGRBs and SLSNe \citep{Blanchard16,Lunnan15}, with a range of sizes, $r_e\approx 0.7-5.6$~kpc. Due to the larger FRB host galaxy sizes among the transient populations, the differences in offsets becomes less significant when normalized by the size of their host galaxies: the host-normalized offsets of FRBs has a median of \medrefoff~$r_e$ (68\% interval of \qrefoff), and are only statistically distinct from Ca-rich transients. Finally, FRBs appear to occur at slightly larger host-normalized offsets than expected given an exponential disk profile. 

In terms of their host galaxy rest-frame optical and NIR light distributions, the FRBs are on moderately fainter regions of their host galaxies (median \FF=\medFF). These distributions serve as proxies for the distribution of stellar mass in their galaxies, and older, moderately massive to low-mass stars, respectively. As a population, FRBs are once again statistically distinct from LGRBs and SLSNe which on average occur on the brighter regions of their host galaxies \citep{Fruchter06,Blanchard16}. The locations of FRBs are also distinct from SGRBs which are very weakly correlated with stellar mass, a consequence of their compact object progenitors which experience kicks and moderate delay times \citep{Fong10,Fong13}. Notably, unlike SGRBs, {\it no} FRBs in our sample occur on the faintest regions of their host galaxies (tempered by the FRB localization errors and the small sample). 

The locations of FRBs are consistent with the {\it radial} distribution of their host rest-frame optical light (fraction of enclosed light), and are indistinguishable from the locations of CCSNe in this regard (c.f., \citealt{Audcent-Ross20}). Indeed, the fact that the locations of FRBs trace the 1:1 distribution of the radial distribution of their host light, coupled with the fact that their local stellar mass surface densities are representative of their global host galaxy values, is indicative that their locations are consistent with the stellar mass within their host galaxies. We further find that while two FRBs have elevated local star formation rate densities compared to their global host values, as a population we do not find any clear correlation between FRBs and regions of elevated local star formation rate densities.

In general, the host galaxies of known repeating FRBs tend to have bluer colors, lower stellar masses, and higher star forming rates than those of apparent non-repeating FRBs \citep{Bhandari20a,Heintz20}. This is most saliently highlighted in the star-forming low-mass host galaxy of the repeating FRB\,121102 \citep{Chatterjee17}, which is an outlier in most host galaxy properties. Here, we find that, in terms of the IR distributions, the three known repeaters in this sample (FRBs\,121102, 180916 and 190711) span the full range of offsets (physical and host-normalized) populated by apparent non-repeaters, as well as fractional flux and enclosed flux. While there do not appear to be any obvious trends in these properties between known repeaters and apparent non-repeaters, we caution that the sample sizes considered here are small.

Overall, the locations of FRBs support the picture that if there is one dominant progenitor population, that they do not originate from massive stars which are stripped of H and/or He (the progenitors of engine-driven SNe such as LGRBs and H-poor SLSNe). We further find that their locations are inconsistent with compact object progenitors which experienced kicks or long delay times
transients from significantly older stellar progenitors (SGRBs, Ca-rich transients). However, given the size of the current sample, it is still possible that a fraction of FRBs originate from one of these alternative progenitor channels. These conclusions overall support previous results based on host stellar population properties that LGRB/SLSNe progenitors are not significant contributors \citep{Heintz20,Li19,Bhandari20a, Bochenek20}. Furthermore, we cannot differentiate the population of FRBs from CCSNe or Type Ia SNe based on their locations, although FRBs do not clearly trace either of these populations in every measured quantity. Therefore, we find it less plausible that the main progenitor channel of FRBs are compact object progenitors such as neutron star mergers or neutron star-black hole mergers, although progenitors which invoke white dwarfs (e.g., accretion-induced collapse of a WD to a NS), which are expected to resemble the properties of Type Ia SNe \citep{Margalit19} could still play a role.

It should be noted, however, that there may be selection effects at play in the observed locations of FRBs.
For example, FRBs in dense regions of galaxies such as near the center or inside a spiral arm may be more difficult to detect. As they interact with a denser medium like a star-forming region, the signal is dispersed by the local environment and the DM could exceed the limit of detection \citep{james21}, i.e.\  
high DM smears the signal leading to a lower S/N. 
The majority of FRBs presented here were derived
from the CRAFT experiment on the ASKAP telescope.
That survey has performed searches allowing 
for bursts with DM$ >1000$~pc cm$^{-3}$ and have detected several to date \citep{Shannon18}. 
Furthermore, the smaller sample of 
well-localized events, including those 
presented here,
follow the predicted Macquart relation \citep{Macquart20}. 
These searches have not detected any bursts with DM greatly in excess of 
the Macquart relation, and models of the 
intrinsic host DM distribution indicate a 
median value of 
$\approx 150$~pc~cm$^{-3}$
\citep{Macquart20, james21}. 
Therefore, current expectation is that there is not
a large sample bursts with high host-DM 
missing from the sample.
However, this is an important effect to consider given the constraints locations provide for progenitor channels and, analysis
of future samples will need to consider 
further the implications of DM smearing.

It also is possible that scattering will be induced by local material increasing the 
width of the burst so that it is too faint to be detected or cause it to be falsely rejected by search algorithms. However, the scattering measure is also known to vary considerably, and this variation does not seem to correlate with DM \citep{Qui20}. 
Therefore, we currently consider this a less important bias
relative to DM smearing.

\subsection{The Association of FRB Locations with Spiral Arm Structure}

In addition to precise location information, the deep {\it HST} imaging presented here also enhances low surface brightness features and morphological structure. In particular, previous {\it HST} studies of two galaxies, those of FRBs\,190608 and 180916, demonstrate that they both exhibit complex spiral arm structure \citep{Chittidi20,tendulkar20}. Spiral structure was also apparent in ground-based imaging for the host galaxy of FRB\,191001, and supported by extended, continuum radio emission indicative of star formation \citep{Bhandari20b}. Here, we find an {\it additional} two FRB hosts with clear spiral arm structure; those of FRBs\,180924 and 190714 (Figure~\ref{fig:galfit_residuals}), and further uncover a bar feature in the host galaxy of FRB\,180924. With the exception of FRB\,180916, all FRB spiral-arm hosts are associated with apparent non-repeaters. The two remaining known repeaters in the sample are FRBs\,121102 and 190711; the former originates from a low-luminosity host, while the latter originates in a host at the high redshift end of our sample. Thus we do not consider the non-detection of spiral features from these hosts to be constraining or informative. 

Overall the prevalence of clear spiral structure (5/8, or $\approx 60\%$ in our sample) is consistent with the observed galaxy population \citep{Willett13}. Furthermore, despite the larger offsets of FRBs, we find that the locations of {\it all} well-localized FRBs with hosts that exhibit spiral structure are consistent with major spiral arm features. It is important to note that the IR light profile is dominated by red supergiants, AGB stars, and low-mass stars, as opposed to young, massive O- and B-stars seen in H$\alpha$ and UV imaging. In particular, in accordance with the density wave theory of spiral structure, the IR spiral arms generally spatially lag the H$\alpha$ light \citep{Pour-Imani16}, although significant enhancement in star formation in the vicinity of the IR spiral arms is expected \citep{Seigar02}. The signal-to-noise of the FRB UV images prevent such a constraint for FRB hosts.

The locations of transients with respect to spiral arm features, as well as offsets from regions of peak brightness within the spiral arms, can serve as a major clue for their progenitors. In particular, the offset from bright peaks serves as a proxy of the spatial drift from birth to explosion site, and can set a timescale for the lifetime of the progenitor. Indeed, SNe exhibit distributions of offsets from the peak of their spiral arms in accordance with their progenitor age, with stripped-envelope SNe (Type Ib/c) having smaller offsets from the peak than Type II or Type Ia SNe \citep{Aramyan16}. If all FRBs originated from young magnetars, it is expected that their positions would generally correlate with the UV spiral arms of their hosts, and at small offsets from star-forming features (c.f., the distribution of Galactic magnetars; \citealt{Olausen14}). However, we find that while the FRB positions are consistent with spiral features, they are not on the brightest part of the spiral arms. Indeed, UV and H$\alpha$ studies of the known repeating FRBs\,121102 and 180916 found clear offsets from the closest star-forming features of $\approx 250$~pc \citep{Bassa17,tendulkar20}. This is also in agreement with the results by \citet{Chittidi20}, who found from detailed analysis of the UV imaging of FRB\,190608 that the FRB did not prefer the most active star-forming region in the galaxy. 

Taken together, this supports a picture that FRBs do not originate from the youngest, most massive stars, in concert with previous, comparative results with other transients \citep{Li19,Heintz20,Bhandari20a}. We also find that FRBs do not appear to reside in the inner bulges of their host galaxies, which are generally dominated by older, higher-metallicity stars in comparison to the spiral arms \citep{Peletier96}. 
It is further worth noting that the main selection effect at play in FRB discovery is the difficulty of detecting highly-scattered FRB signals, where the signal is temporally broadened by multipath propagation in a dense, turbulent medium. Since such sites are preferentially associated with star formation, one might naively expect there to be additional observational challenges in detecting FRBs in spiral arms where the chance of the FRB sightline intersecting an enhanced region of turbulence is higher. However, the precise effects of discovering FRBs with respect to morphological structure is not well-quantified.

\subsection{Luminosity Limits on Alternative Host Candidates}

Finally, we remark on the presence of fainter, alternative host galaxy candidates at the positions of the FRBs. This question is in part motivated by the low-luminosity host galaxy of FRB\,121102 \citep{Tendulkar17}, which, coupled with the remaining FRB hosts suggests a broad host galaxy luminosity function spanning the full range of galaxies \citep{Heintz20}. Here we have explored the presence of both satellite galaxies at the same redshift as the putative host, and background galaxies at higher redshifts. The relatively low redshift range of the population examined here, $0.03 \lesssim z \lesssim 0.522$, enables deep constraints even in the presence of strong morphological features. In both scenarios, we find it unlikely that the FRBs originated from an underlying galaxy. The exceptions are FRBs\,190714 and 191001: in the former case, a galaxy of equal luminosity to FRB\,121102 would approach the redshift limit, while in the latter case, the high DM allows a host with $\approx 8$ times the luminosity of FRB\,121102, albeit still on the faint-end slope of the galaxy luminosity function ($\approx 8 \times 10^{8}~L_{\odot}$).

\section{Summary \& Conclusions}
\label{sec:conc}

In this paper we used high-resolution {\it HST} imaging to perform a detailed study on the locations of 8 FRBs and their environments, 6 of which are newly presented here. We used these data to place constraints on the spatial distributions (physical and host-normalized), in support of previous works based on ground-based imaging. We find a median host-normalized offset of \medrefoff (\qrefoff; 68\% interval), and overall a distribution that lies between the more centrally concentrated LGRBs and SLSNe, and the extended SGRBs and Ca-rich transients. We also determine the distribution of FRBs with respect to their IR (rest-frame optical) host galaxy light (fractional flux, and radial distribution), showing that FRBs are consistent with tracing the stellar mass distribution of their host galaxies.

The sensitivity of {\it HST} additionally enables constraints on possible alternative host galaxy candidates; we find it improbable that there exists a satellite or background galaxy at the FRB locations, strengthening the associations with the brighter, putative hosts identified in ground-based imaging for this sample. We explore the FRB site properties in terms of star formation rate (near-UV) and stellar mass (IR) surface densities, finding that the locations are not particularly enhanced in either property compared to the global values of their hosts (although few measurements exist for the star formation rate densities). Finally, we find that 5/8 FRB host galaxies in the sample have spiral arm features, and that these FRBs are consistent with the locations of those spiral arms (albeit inconsistent with locations on the {\it brightest} peaks of these spiral arms). If there is a dominant progenitor population among this tested distribution, we thus do not find strong support for a connection to the most massive (stripped-envelope) stars, or events which require kicks and long delay times.

The promise of sub-arcsecond localized FRBs in solving the progenitor question is being realized, in part, with the first population studies of their local and host galaxy environments. Such precisely-localized FRBs are and will continue to be detected at growing rates. As the number of secure associations continues to increase, we will be able to make significant progress toward understanding their progenitors, as well as connecting the properties of FRBs to those of their host galaxies. The current sample of well-localized FRBs is admittedly small, much less those with high-resolution imaging. Moreover, the current sample with secure host galaxies is subject to various selection biases which have been mentioned throughout this paper. 

Looking toward the future, upcoming FRB experiments and upgrades to existing ones will deliver larger, more uniform samples of sub-arcsecond localized FRBs, which importantly will push beyond the current DM or redshift horizons. Equipped with a large sample of FRBs with high-resolution imaging, we will be able to identify trends between the locations of FRBs in known repeaters and apparent non-repeaters, in host galaxies of different morphological types (e.g., prevalence of spiral sub-structure, star-forming vs. quiescent), and make more statistically significant statements about their similarities or differences to other transient populations. Matched to the increased sensitivities of discovery experiments, we will also explore evolution of the local properties of FRBs with redshift. All of these studies will provide important clues to their origins. Larger samples will also enable tighter constraints to be placed on local contributions to the dispersion measure in DM and IGM studies, optimizing the use of FRBs as a cosmological probe. Finally, {\it HST} and soon {\it JWST} will also aid in our understanding of whether FRBs originate from a single, dominant progenitor channel or multiple contributing channels, a central question in FRBs.

\software{Photutils \citep{photutils}; GALFIT \citep{Peng02}; Source Extractor \citep{SExtractor}; IRAF \citep{Tody86}; NumPy \citep{numpy}; Astropy \citep{astropy:2018}; Matplotlib \citep{matplotlib}; SciPy \citep{scipy}.}

\facilities{HST:WFC3}

\section*{Acknowledgements}

We acknowledge Nia Imara, Clancy W. James, and Ben Margalit for helpful discussions. A.G.M. acknowledges support by the National Science Foundation Graduate Research Fellowship under Grant No. 1842400. The Fast and Fortunate for FRB Follow-up team acknowledges support from NSF grants AST-1911140 and AST-1910471. W.F. acknowledges support by the National Science Foundation under grant Nos. AST-1814782 and AST-190935. N.T. acknowledges support by FONDECYT grant 11191217. C.K.D. acknowledges the support of the CSIRO Postgraduate Scholarship - Astronomy and Space (47417). A.T.D. is the recipient of an ARC Future Fellowship (FT150100415). This research is based on observations made with the NASA/ESA Hubble Space Telescope obtained from the Space Telescope Science Institute, which is operated by the Association of Universities for Research in Astronomy, Inc., under NASA contract NAS 5–26555. These observations are associated with programs \#15878, 16080, 14890 and 16072. Support for Program numbers 15878 and 16080 were provided through a grant from the STScI under NASA contract NAS5- 26555.

\bibliography{hst_refs}

\end{document}